\documentclass[accepted=2018-11-29]{quantumarticle}
\pdfoutput=1
\usepackage[utf8]{inputenc}
\usepackage[numbers,sort&compress]{natbib}
\usepackage{amsmath,amssymb,amsfonts,bbm,graphicx}

\newcommand{\ket}[1]{\vert #1 \rangle}

\newcommand{\id}{\mathbbm{1}}
\newcommand{\qunr}[1]{ \widetilde{\qfi}_{\mathsf{unr}, #1}}

\newcommand{\phd}{{\vphantom{\dag}}}
\def\Tr{\hbox{Tr}}
\newcommand{\qfi}{\mathcal{Q}}
\begin{document}
\title{Restoring Heisenberg scaling in noisy quantum metrology by monitoring the environment}
\author{Francesco Albarelli}
\affiliation{Quantum Technology Lab, Dipartimento di Fisica ``Aldo Pontremoli'', Università degli Studi di Milano, IT-20133, Milan, Italy}
\affiliation{Department of Physics, University of Warwick, Coventry CV4 7AL, United Kingdom}
\orcid{0000-0001-5775-168X}
\author{Matteo A. C. Rossi}
\affiliation{Quantum Technology Lab, Dipartimento di Fisica ``Aldo Pontremoli'', Università degli Studi di Milano, IT-20133, Milan, Italy}
\affiliation{QTF Centre of Excellence, Turku Centre for Quantum Physics, Department of Physics and Astronomy, University of Turku, FI-20014 Turun Yliopisto, Finland}
\orcid{0000-0003-4665-9284}
\author{Dario Tamascelli}
\affiliation{Quantum Technology Lab, Dipartimento di Fisica ``Aldo Pontremoli'', Università degli Studi di Milano, IT-20133, Milan, Italy}
\orcid{0000-0001-6575-4469}
\author{Marco G. Genoni}
\affiliation{Quantum Technology Lab, Dipartimento di Fisica ``Aldo Pontremoli'', Università degli Studi di Milano, IT-20133, Milan, Italy}
\orcid{0000-0001-7270-4742}
\email{marco.genoni@fisica.unimi.it}
%
%
\begin{abstract}
We study quantum frequency estimation for $N$ qubits subject to independent Markovian noise via strategies based on time-continuous monitoring of the environment.
Both physical intuition and the extended convexity of the quantum Fisher information (QFI) suggest that these strategies are more effective than the standard ones based on the measurement of the unconditional state after the noisy evolution.
Here we focus on initial GHZ states subject to parallel or transverse noise. For parallel, i.e., dephasing noise, we show that perfectly efficient time-continuous photodetection allows us to recover the unitary (noiseless) QFI, and hence obtain Heisenberg scaling for every value of the monitoring time.
For finite detection efficiency, one falls back to noisy standard quantum limit scaling, but with a constant enhancement due to an effective reduced dephasing.
In the transverse noise case, Heisenberg scaling is recovered for perfectly efficient detectors, and we find that both homodyne and photodetection-based strategies are optimal.
For finite detector efficiency, our numerical simulations show that, as expected, an enhancement can be observed, but we cannot give any conclusive statement regarding the scaling.
We finally describe in detail the stable and compact numerical algorithm that we have developed in order to evaluate the precision of such time-continuous estimation strategies, and that may find application in other quantum metrology schemes.
\end{abstract}
\maketitle
\section{Introduction}
Quantum metrology is one of the most promising fields within the realm of quantum technologies, with applications ranging from spectroscopy and magnetometry to interferometry and gravitational waves detection~\cite{Caves1981,Holland1993,Bollinger1996,McKenzie2002}.
While $N$ uncorrelated (\emph{classical}) probe states lead to an estimation precision scaling as $1/\sqrt{N}$, typically referred to as \emph{standard quantum limit} (SQL), quantum probes made of $N$ entangled particles allow to design schemes with precision scaling as $1/N$,  reaching the so-called \emph{Heisenberg limit} (HL)~\cite{GiovannettiNatPhot}.

The above result relies on the assumption of noiseless unitary dynamics, but the unavoidable interaction with the surrounding environment can have dramatic consequences on the performances of these protocols. When the dynamics is described by a dephasing dynamical semigroup, the estimation precision is bounded to follow a SQL scaling, and thus only a constant enhancement can be obtained by exploiting quantum resources~\cite{Huelga97,EscherNatPhys,KolodynskyNatComm,Koodynski2013}.
Several attempts to circumvent these no-go theorems and obtain a superclassical scaling in the presence of noise have been pursued, exploiting time-inhomogeneous dynamics~\cite{Matsuzaki2011,Chin12,Smirne16,Haase2017,Gorecka2017}, noise with a particular geometry~\cite{Chaves2013,Brask15}, dynamical decoupling~\cite{Sekatski2015a}, and quantum error-correction protocols (or, more generally, the possibility to implement control Hamiltonians)~\cite{Kessler14,Arrad14,Dur14,Plenio2016,Gefen2016,Layden2017,Sekatski2017metrologyfulland,Matsuzaki2017,Zhou2018}.

Our goal is to attack the problem of noisy quantum metrology by exploiting time-continuous monitoring~\cite{WisemanMilburn,SteckJacobs}.
Time-continuous measurements have been often studied as a tool for quantum parameter/state estimation~\cite{Mabuchi1996,Verstraete2001,Gambetta2001,Chase2009,Ralph2011a,Six2015,Cortez2017,Ralph2017}, with a particular focus on classical time-dependent signals~\cite{Tsang2009b,Tsang2011,Ng2016}.
More recently, methods to calculate classical and quantum Cramér-Rao bounds for parameter estimation via time-continuous monitoring have been proposed~\cite{Guta2011,GammelmarkCRB,GammelmarkQCRB,Macieszczak2016,Genoni2017}, and put into action~\cite{KiilerichPC,Kiilerich2015a,KiilerichHomodyne,Albarelli2017a}.

In this paper, we apply such techniques to the problem of frequency estimation for an ensemble of two-level atoms (qubits), each subjected to independent Markovian noise.
In particular, we consider estimation strategies based on time-continuous (\emph{weak}) measurements of each qubit via the monitoring of the corresponding environment, and a final (\emph{strong}) measurement on the conditional state of the $N$ qubits.
Few similar attempts in this sense have already been discussed in the literature, but they all present substantial differences compared to our approach.
For instance, in~\cite{Geremia2003,Molmer2004,Albarelli2017a} the same problem of frequency estimation (magnetometry) is addressed, but in the presence of collective transverse noise. There, a single \emph{environment} collectively interacts with all the qubits and the corresponding noise is not detrimental for the Heisenberg scaling; time-continuous monitoring is however extremely promising as, thanks to the corresponding back-action on the system, a Heisenberg-limited precision can be achieved even by starting the dynamics with an uncorrelated spin-coherent state.
An estimation problem more similar to ours is presented in~\cite{Catana2014}, where independent environments interact with each probe, but the analysis is restricted to a single (discrete) step of the dynamics and to interaction Hamiltonians commuting with the generator of the phase rotation (i.e., in the presence of pure dephasing only).

Here we consider a proper continuous evolution, described by a Markovian master equation in the Lindblad form, leading to a dynamics satisfying the semigroup property. We focus in particular on independent noise, either parallel or transverse to the generator of the phase rotation to be estimated.
In the former case, which physically corresponds to pure dephasing, the unconditional dynamics leads to a standard quantum limited precision, even for an infinitesimal amount of noise~\cite{KolodynskyNatComm}.
In the latter, it was shown that, by optimizing over the evolution time, it is possible to restore a super-classical scaling between SQL and HL~\cite{Chaves2013,Brask15}.

Our goal is to study whether in both cases time-continuous monitoring will allow for the restoration of the HL, and to analyze in detail the effect of the monitoring efficiency on the performance of the estimation schemes.
First, we will derive the ultimate limit on estimation precision, optimizing over the most general measurements on the joint degrees of freedom of system and environment.
We will then restrict ourselves to the strategies briefly described above, focusing on time-continuous photodetection and homodyne-detection.

To achieve those aims, we develop a stable and compact numerical algorithm that allows us to calculate the effective quantum Fisher information characterizing this kind of measurement strategy, and that will find application in quantum metrology problems beyond the ones considered in this paper.

The manuscript is structured as follows: In Sec.~\ref{s:qest} we introduce the basic concepts of quantum estimation theory, with a particular focus on measurement strategies based on time-continuous measurements.
In Sec.~\ref{s:results} we first introduce the problem of noisy quantum frequency estimation and then we present our original results for parallel and transverse Markovian noise.
The following sections are devoted to describing the methods exploited to obtain such results.
In Sec.~\ref{s:UQFI} we show how to evaluate analytically the ultimate limits posed by quantum mechanics for strategies optimizing over all the the possible global measurements on system and environment.
In Sec.~\ref{s:algorithm} we present in detail the numerical algorithm we have developed for the calculation of the effective quantum Fisher information, pertaining to strategies based on time-continuous monitoring of the environment.
Sec.~\ref{s:conclusion} concludes the paper with some final remarks and discussion of possible future directions.

\section{Quantum estimation via time-continuous measurements} \label{s:qest}
A classical estimation problem is typically described in terms of a conditional probability $p(x|\omega)$ of obtaining the measurement outcome $x$, given the value of the parameter $\omega$
that one wants to estimate.
The classical Cram\'er-Rao bound states that the precision of any unbiased estimator is lower bounded as
\begin{equation}
\delta \omega \geq \frac{1}{\sqrt{M \mathcal{F}[p(x|\omega)]}} \,,
\end{equation}
where $M$ is the number of measurements performed, $\mathcal{F}[p(x|\omega)] =
\mathbbm{E}_p[(\partial_\omega \ln p(x|\omega) )^2]$ is the classical Fisher information,
and $\mathbbm{E}_p[\cdot ]$ denotes the average over the probability distribution $p(x|\omega)$.

In the quantum setting, the probability is obtained via the Born rule, $p(x|\omega)=\Tr[\varrho_\omega \Pi_x]$,
where $\varrho_\omega$ is a family of quantum states parametrized by $\omega$, and $\Pi_x$ is an element of the positive-operator valued measure (POVM) describing the measuring process.
It is then possible to optimize over
all the possible POVMs, obtaining the quantum Cram\'er-Rao inequality \cite{CavesBraunstein}
\begin{align}
\delta \omega \geq \frac{1}{\sqrt{M \mathcal{F}[p(x|\omega)]}} \geq \frac{1}{\sqrt{M \qfi[\varrho_\omega]}}\,, \label{eq:QCRB}
\end{align}
where
\begin{align}
\qfi[\varrho_\omega] =
\lim_{\epsilon \rightarrow 0} \frac{ 8 \left(1 - F \left[ \varrho_\omega, \varrho_{\omega + \epsilon}  \right] \right) }{ \epsilon^2 }
\end{align}
is the quantum Fisher information (QFI), expressed in terms of the fidelity between quantum states $F \left[\varrho_1,\varrho_2 \right] = \Tr \left[ \sqrt{\sqrt{\varrho_1} \varrho_2 \sqrt{\varrho_1} } \right] $.
The QFI $\qfi[\varrho_\omega]$ clearly depends only on the quantum statistical model, namely the $\omega$-parametrized family of quantum states $\varrho_\omega$, and one can always find the optimal POVM saturating the bound., i.e., such that $\mathcal{F}[p(x|\omega)]=\qfi[\varrho_\omega]$.

In several quantum parameter estimation problems, including the case of standard frequency estimation, $\varrho_\omega$ corresponds to the \emph{unconditional} state evolved according to a Markovian master equation of the form
\begin{equation}
\frac{d\varrho}{dt} = \mathcal{L}_\omega \varrho =  -i [ \hat{H}_\omega,\varrho] + \sum_{j} \mathcal{D}[\hat{c}_j] \varrho \,, \label{eq:Markov}
\end{equation}
where the parameter is encoded in the Hamiltonian $\hat{H}_\omega$,
the operators $\hat{c}_j$ denote different independent noisy channels, and we have defined the superoperator
$\mathcal{D}[\hat{A}] \varrho = \hat{A}\varrho \hat{A}^\dag - (\hat{A}^\dag \hat{A} \varrho + \varrho \hat{A}^\dag \hat{A})/2$.

The master equation above can be obtained assuming that the system is interacting with a
sequence of input operators $\hat{a}^{(j)}_\mathsf{in}(t)$, one for each noise operator $\hat{c}_j$, that satisfy the bosonic commutation relation $[\hat{a}^{(j)}_\mathsf{in}(t),\hat{a}^{(k) \dag}_\mathsf{in}(t')]=\delta_{jk}\delta(t-t')$.
Some information on the parameter $\omega$ is then contained in the corresponding output operators $\hat{a}^{(j)}_\mathsf{out}(t)$, i.e., the environmental modes, just after the interaction with the system.
One can imagine to perform a joint measurement on all output modes and the quantum system itself.
The ultimate limit on the estimation of $\omega$ that takes into account all these strategies based on measurements of system and output, is quantified by the QFI introduced in~\cite{GammelmarkQCRB}.
It depends only on the initial state of the system and on the superoperator $\mathcal{L}_\omega$ describing the evolution.
In general, it can be evaluated as
\begin{align}
\label{eq:ultimQFI}
	\mathcal{\overline{Q}}_{\mathcal{L}_\omega} = 4 \partial_{\omega_1} \partial_{\omega_2} \log  \left| \Tr[\bar\varrho ]\right| \big|_{\omega_1 = \omega_2 = \omega}\,\, ,
\end{align}
where, in the case we are considering (i.e., a Hamiltonian parameter), the operator $\overline\varrho$ is the one solving the generalized master equation
\begin{align}
\label{eq:MolmerGenME}
\frac{d\bar\varrho}{dt} &=  \mathcal{L}_{\omega_1,\omega_2} \bar\varrho \notag \\
&=  -i \left( \hat{H}_{\omega_1}\bar\varrho -\bar\varrho \hat{H}_{\omega_2} \right) + \sum_{j} \mathcal{D}[\hat{c}_j] \bar\varrho \,.
\end{align}

This QFI corresponds to an optimization of the classical FI over all the possible measurement strategies described above, including the possibility of performing non-separable (entangled) measurements over the system and all the output modes at different times.

However, one can restrict to more feasible strategies, where the outputs are sequentially measured continuously in time, and a final \emph{strong} measurement is performed on the conditional state of the system.
Depending on the type of measurement performed on the outputs, one obtains different stochastic master equations for the conditional state of the system.
In the following we will focus on the two paradigmatic cases of photodetection (PD) and homodyne detection (HD).

In the first case, assuming time-continuous PD on each output with efficiency $\eta_j$, the evolution is described by the stochastic master equation~\cite{WisemanMilburn}
\begin{align}
d\varrho^{(c)} &= - i [\hat{H}_{\omega} , \varrho^{(c)}]\,dt + \sum_j (1-\eta_j) \mathcal{D}[\hat{c}_j] \varrho^{(c)} \,dt \notag \\
& -\frac{\eta_j}{2} (\hat{c}_j^\dag \hat{c}^\phd_j \varrho^{(c)} +\varrho^{(c)} \hat{c}_j^\dag \hat{c}^\phd_j ) \,dt + \eta_j \Tr[\varrho^{(c)} \hat{c}_j^\dag \hat{c}^\phd_j ]\varrho^{(c)} \,dt \notag \\
& + \left( \frac{\hat{c}^\phd_j \varrho^{(c)} \hat{c}_j^\dag}{\Tr[\varrho^{(c)} \hat{c}_j^\dag \hat{c}^\phd_j ] } - \varrho^{(c)} \right)dN_j  \,, \label{eq:photoSME}
\end{align}
where $dN_j$ denote Poisson increments taking value $0$ (no-click event) or $1$ (detector click event), and having average value $\mathbbm{E}[dN_j] = \eta_j \Tr[\hat{c}_j^\dag \hat{c}^\phd_j \varrho^{(c)}]\,dt$.

Likewise, for time-continuous HD on each output, the stochastic master equation reads~\cite{WisemanMilburn,Rouchon2015}
\begin{align}
d\varrho^{(c)} ={}& - i [\hat{H}_{\omega} , \varrho^{(c)}]\,dt + \sum_j \mathcal{D}[\hat{c}_j] \varrho^{(c)} \,dt \notag \\
& + \sum_j \sqrt{\eta}_j  \mathcal{H}[\hat{c}_j]\varrho^{(c)} \, dw_j \:,\label{eq:homoSME}
\end{align}
where  $\mathcal{H}[\hat{c}]\varrho = \hat{c} \varrho + \varrho \hat{c}^\dag - \Tr[\varrho (\hat{c} +\hat{c}^\dag)]\varrho$, and
 $dw_j = dy_j - \sqrt{\eta}_j \Tr[\varrho^{(c)} (\hat{c}^\phd_j + \hat{c}_j^\dag )]$ represents a standard Wiener increment (s.t.
$dw_j dw_k = \delta_{jk} dt$), operationally corresponding to the difference between the measurement
result $dy_j$ and the rescaled average value of the operator $(\hat{c}^\phd_j + \hat{c}_j^\dag)$.

In general, different measurement strategies mathematically correspond to different possible unravellings of the Markovian master equation: any quantum state, solution of Eq.~\eqref{eq:Markov} at a certain time $t$, can be written as $\varrho_\mathsf{unc} = \sum_\mathsf{traj} p_\mathsf{traj} \varrho^{(c)}$, where $p_\mathsf{traj}$ is the probability of a certain trajectory leading to the conditional state $\varrho^{(c)}$, and where the sum is replaced by an integral in the case of time-continuous measurements with a continuous spectrum (e.g. homodyne).

One can then define an effective QFI, which poses the ultimate bounds for these kinds of estimation strategies~\cite{Catana2014,Ng2016,Albarelli2017a}, and that depends both on the specific unravelling and on the monitoring efficiencies $\eta_j$:
\begin{align}
\widetilde{\qfi}_{\mathsf{unr},\eta_j}  = \mathcal{F}[p_\mathsf{traj}] + \sum_\mathsf{traj} p_\mathsf{traj} \qfi[\varrho^{(c)}] \,. \label{eq:effQFI}
\end{align}
As it is apparent from the formula, $\widetilde{\qfi}_{\mathsf{unr},\eta} $ is equal to the sum of the classical Fisher information of the trajectories probability distribution $\mathcal{F}[p_\mathsf{traj}]=\sum_{\mathsf{traj}} \frac{ (\partial_\omega p_\mathsf{traj} )^2 }{p_\mathsf{traj}}$, plus the average of the QFIs of the corresponding conditional states $\qfi[\varrho^{(c)}]$.
The first term quantifies the amount of information obtained from the measurement of the output modes after the interaction with the system, while the second term quantifies the amount of information obtained from the final strong measurement on the conditional states of the system itself. We point out that an analogous figure of merit has been recognized as the appropriate one for metrology with post-selection~\cite{Combes2014,Zhang2015,Alves2015}.

A method for the calculation of $\mathcal{F}[p_\mathsf{traj}]$ has been firstly proposed in \cite{GammelmarkCRB} for generic quantum states, and then in~\cite{Genoni2017} for continuous-variable Gaussian states.
In Sec.~\ref{s:algorithm}, we describe in detail a more compact and stable numerical algorithm for the calculation of the terms present in Eq.~\eqref{eq:effQFI}, for both homodyne and photodetection measurements.

It is reasonable to expect that, thanks to the information obtained from the time-continuous monitoring, and thanks to the, typically beneficial, effects of quantum measurements on the conditional states, one will obtain more information on the parameter via these strategies, than by solely measuring the unconditional state solution of the Markovian master equation \eqref{eq:Markov}.
This intuition is confirmed by the extended convexity property of the QFI, recently proved in~\cite{Alipour2015,Ng2016}.
In fact, the following chain of inequalities holds:
\begin{align}\label{eq:QFIineq}
\qfi[\varrho_\mathsf{unc}] &\leq \qunr{\eta_j}  \leq \overline{\qfi}_\mathcal{L_\omega}  \,.
\end{align}

We also conjecture that, for a fixed unravelling, and for fixed efficiency on all the output channels (i.e., for $\eta_j=\eta \, \forall j$), the effective QFI is monotonic with respect to the efficiency parameter, that is:
\begin{align}\label{eq:QFIineq_conj}
\qunr{\eta}  \leq \qunr{\eta^\prime} \quad \Longleftrightarrow
\quad \eta \leq \eta^\prime \qquad \textrm{(conjecture)}\,.
\end{align}
\section{Quantum frequency estimation in the presence of noise} \label{s:results}
We consider a system of $N$ qubits, described by a Hamiltonian $\hat{H}_\omega = (\omega/2) \sum_{j=1}^N \sigma_z^{(j)}$, where $\omega$ is the unknown frequency to be estimated, and $\sigma_z^{(j)}$ is the Pauli-z operator acting on the $j$-th qubit.
The system is interacting also with a Markovian environment, so that the evolution is described by the master equation
\begin{align}
\frac{d\varrho}{dt} &= \mathcal{L}_\omega \varrho =  -i [ \hat{H}_\omega,\varrho] + \frac\kappa{2} \sum_{j=1}^N \mathcal{D}[\sigma_\alpha^{(j)}] \varrho \,, \label{eq:MarkovFreq}
\end{align}
where $\alpha \in \{ x , z \} $, i.e., we only consider noise parallel or transverse to the Hamiltonian.
The master equation~\eqref{eq:MarkovFreq} can be easily mapped to Eq.~\eqref{eq:Markov} by considering the noise operators $\hat{c}_j = \sqrt{\kappa/2} \sigma_\alpha^{(j)}$.
In what follows we will study all the quantities described in the previous section in order to assess the estimation of the parameter $\omega$.

In quantum frequency estimation strategies, one considers the number of qubits $N$ and the total time of the experiment $T$ as the resources of the protocol.
The quantum Cram\'er-Rao bound~\eqref{eq:QCRB} is then more efficiently rewritten as
\begin{align}
\delta \omega \sqrt{T} \geq \frac{1}{\sqrt{\qfi/t}} \geq \frac{1}{\sqrt{ \max_t [\qfi/t ]}} \label{eq:QCRBomega}
\end{align}
where $t=T/M$ corresponds to the duration of each round, over which one can perform a further optimization, and where $\qfi$ corresponds here to the proper QFI charaterizing the particular estimation strategy considered.

In the rest of the manuscript we are restricting to initial entangled GHZ states $|\psi_\mathsf{GHZ}\rangle = ( |0\rangle^{\otimes N} + |1\rangle^{\otimes N})/\sqrt{2}$.
It is well known that in the noiseless case, i.e., for $\kappa=0$, the corresponding QFI is Heisenberg limited, $\qfi_\mathsf{HL}= N^2 t^2$. This leads to a quadratic enhancement w.r.t. the ``standard quantum limited'' QFI, $\qfi_\mathsf{SQL} = N t^2$ (obtained in the case of a factorized \emph{coherent-spin} initial state $|\psi_\mathsf{coh}\rangle = [(|0\rangle + |1\rangle)/\sqrt{2}]^{\otimes N}$).

In what follows, we will consider the noisy case ($\kappa>0$); the generic QFI $\qfi$ in Eq.~\eqref{eq:QCRBomega} will correspond to either:
i) the QFI of the unconditional state $\qfi[\varrho_\mathsf{unc}]$ corresponding to the master equation~\eqref{eq:MarkovFreq};
ii) the ultimate QFI $\overline{\qfi}_{\mathcal{L}_\omega}$ obtained optimizing over all the possible measurements on system and environmental outputs;
iii) the effective QFI $\widetilde{\qfi}_{\mathsf{unr},\eta}$ corresponding to a specific time-continuous (sequential) measurement of the output modes and a final strong measurement on the conditional state of the system.
In particular we are will focus on time-continuous photodetection and homodyne detection, with measurement efficiency $\eta$ and we are labelling the respective effective QFIs as $\widetilde{\qfi}_\mathsf{pd,\eta}$ and $\widetilde{\qfi}_\mathsf{hom,\eta}$.
It is important to remark that, given the master equation~\eqref{eq:MarkovFreq}, one is assuming that $N$ different (homodyne or photo-) detectors are monitoring the environment of each qubit. We will however find instances where this assumptions may be relaxed.

\begin{figure}[t]
\centering
\includegraphics[width=.9\columnwidth]{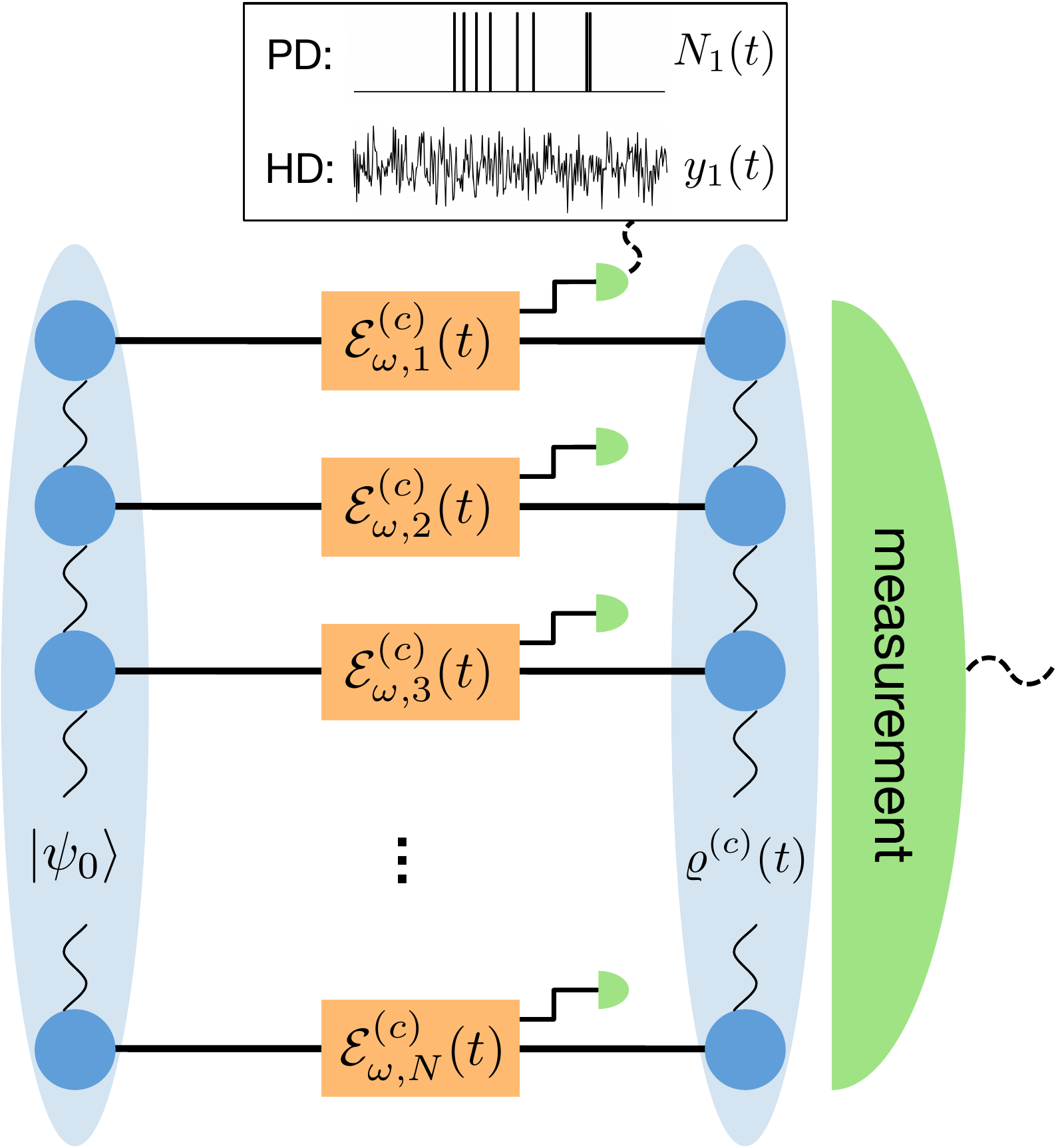}
\caption{Schematic representation of the metrological approach we consider in this paper.
A $N$-qubit (possibly entangled) input state $| \psi_0 \rangle$ interacts with $N$ independent environments which are monitored by $N$ detectors, either by photodetection (PD) or homodyne detection (HD).
In the former case each output is a binary valued detection record $N_j (t)$, in the latter a real valued photo-current $y_j (t)$.
The map $\mathcal{E}^{(c)}_{\omega,j}(t)$ represents the corresponding conditional dynamics, i.e., the solution of the stochastic master equation for the single $j$-th qubit only, either for PD~\eqref{eq:photoSME} or HD~\eqref{eq:homoSME}, and thus depending on the corresponding measurement output $N_j(t)$ or $y_j(t)$.
The $N$-qubit quantum state $\varrho^{(c)}(t)$, conditioned on all the measurement outcomes acquired during the evolution, is further collectively measured at the end of the conditional dynamics.}
\label{fig:diagram}
\end{figure}

In the next subsections we address separately the two different cases of parallel and transverse noise.
For each case we start by reviewing known results for the QFI of the unconditional states; then we will present our original results, regarding the ultimate QFI $\overline{\qfi}_{\mathcal{L}_\omega}$ and effective QFIs $\widetilde{\qfi}_\mathsf{pd,\eta}$ and $\widetilde{\qfi}_\mathsf{hom,\eta}$ (corresponding to the schemes pictorially represented in Fig.~\ref{fig:diagram}), without dwelling into the details of their calculation, which will be left to Secs.~\ref{s:UQFI} and~\ref{s:algorithm}.

 \subsection{Parallel noise}
Parallel noise, corresponding to the master equation~\eqref{eq:MarkovFreq} with $\sigma_\alpha^{(j)}=\sigma_z^{(j)}$, is typically considered the most detrimental noise for frequency estimation since the evolution induces dephasing in the eigenbasis of the Hamiltonian $\hat{H}_\omega$.
For an initial GHZ state, the QFI of the unconditional state can be evaluated analytically~\cite{Huelga97,Alipour2014}, obtaining $\qfi[\varrho_\mathsf{unc}] = N^2 t^2 e^{- 2 \kappa N t}$.\\
By optimizing over the single-shot duration $t$, one obtains that the optimal QFI is standard quantum limited,
\begin{align}
\max_t \left[\frac{\qfi[\varrho_\mathsf{unc}]}{t} \right] = \frac{N}{2 e \kappa}  \,. \label{eq:QFIparUnc}
\end{align}
This result is equivalent to the one obtainable via a factorized initial state.
Only a constant enhancement can in fact be gained, by optimizing over the initial entangled state~\cite{Huelga97}; by doing this one saturates the ultimate noisy bound derived in~\cite{EscherNatPhys,KolodynskyNatComm}, which dictates that, as soon as some parallel noise is present in the dynamics, the ultimate precision is standard quantum limited.

On the other hand, the ultimate QFI $\overline{\qfi}_{\mathcal{L}_\omega}$ can be easily evaluated (see Sec.~\ref{s:UQFI} for details), and it turns out to be equal to the noiseless QFI, i.e.
\begin{align}
\overline{\qfi}_{\mathcal{L}_\omega}^\parallel = N^2 t^2 \,.
\end{align}
This shows that, by measuring also the output modes, it is in principle possible to recover not only a Heisenberg scaling for the error $\delta\omega$, but also the whole information on the parameter.

For both time-continuous homodyne and photodetection, the evolution of an initial GHZ state, under parallel noise and conditioned on the measurement results, is restricted to a two-dimensional Hilbert space.
As a consequence, the results obtained for $N=1$ qubit can be readily used to infer the results for generic $N$ qubits, by simply rescaling the evolution time $t \rightarrow Nt$ (see Sec.~\ref{s:algorithm} for details).
We can thus obtain numerically exact values of the effective QFIs for any value of $N$; in particular we have been able to numerically verify that the strategy based on time-continuous photodetection with perfect efficiency $\eta=1$ is indeed optimal, i.e.
\begin{align}
\widetilde{\qfi}^\parallel_{\mathsf{pd},\eta=1} = \overline{\qfi}_{\mathcal{L}_\omega}^\parallel = N^2 t^2 \,,
\end{align}
showing that the noiseless Heisenberg-limited result can be recovered, without the need to perform complicated (non-local in time) measurement strategies on system and environment.

The physical explanation of this result can be easily understood by studying the action of the two Kraus operators describing the conditional dynamics for initial GHZ states (see Sec.~\ref{s:algorithm} for more details on the Kraus operators).
The \emph{no-jump} evolution is ruled by the infinitesimal Kraus operator $M_0 = \mathbbm{1} - i \hat{H}_\omega \,dt +(\kappa N/4)\mathbbm{1} dt$; as a consequence it easy to check that the GHZ state evolves as in the unitary case, apart from an irrelevant normalization factor that can be ignored.

On the other hand, when a jump occurs due to a photon detected in the output corresponding to one of the $N$ qubits, the only effect of the corresponding Kraus operator, $M_1^{(j)} = \sigma_z^{(j)} \sqrt{(\kappa/2) dt}$, is to induce a relative minus sign between the $|0\rangle^{\otimes N}$ and the $|1\rangle^{\otimes N}$ components of the quantum conditional state, independently on the index $j$ of the corresponding qubit.
In general, after a time $t$ and $m$ detected photons from all the the output channels, the conditional state reads:
 \begin{align}
|\psi_{m\lvert\mathsf{GHZ}}\rangle = \frac{1}{\sqrt{2}}\left( |0\rangle^{\otimes N} + e^{i (N \omega t + m \pi)} |1\rangle^{\otimes N} \right) \,.
 \end{align}

It is important to remark that, thanks to the symmetry of the GHZ state, and given that jumps may occur only one by one~\cite{WisemanMilburn}, it is not necessary to know exactly from which qubit channel the photon has been detected; as a consequence one may obtain the same result by using a single perfect photo-detector monitoring jointly all the $N$ output modes.
Once the number of detected photons $m$ is known, and the corresponding ``GHZ-equivalent'' conditional state is prepared, one can estimate the frequency $\omega$ at the Heisenberg limit.

As soon as the photodetection monitoring is not perfectly efficient (we always consider equal efficiency for all the qubits, $\eta_j = \eta < 1, \,\, \forall j$), the Heisenberg scaling is immediately lost.
Our numerical results clearly show that the effective QFI is equal to $\widetilde{\qfi}^\parallel_{\mathsf{pd},\eta} = N^2 t^2 e^{- 2 \kappa (1-\eta) N t}$, and the optimized QFI reads
\begin{align}
 \max_t \left[\frac{\widetilde{\qfi}^\parallel_{\mathsf{pd},\eta}}{t} \right] = \frac{N}{2 e \kappa (1-\eta)}  \,.
\end{align}

Time-continuous inefficient photodetection thus leads to a constant enhancement, compared to the unconditional/classical case~\eqref{eq:QFIparUnc}, which physically corresponds to have a reduced effective dephasing parameter $\kappa_\mathsf{eff} = \kappa (1-\eta)$.
We remark that also in this case it is not necessary to monitor every output channel separately, and that a single-photo detector can be employed.
Moreover, in this picture, the overall efficiency parameter $\eta$ corresponds to the product between the factual efficiency of the detectors and the fraction of qubits that are effectively monitored.

We also notice that, for every value of $\omega$, all the information is contained in the conditional quantum states: the classical Fisher information $\mathcal{F}[p_\mathsf{traj}]$ is in fact identically equal to zero.
This follows from the unitarity of the Pauli matrices, leading to $\hat{c}_j^\dag \hat{c}^\phd_j = \kappa \mathbbm{1}_2/2$, and thus to a parameter-independent Poisson increment with average value $\mathbbm{E}[dN_j] = (\eta \kappa/2)\, dt$ in  Eq. \eqref{eq:photoSME}.
Nevertheless, we remark that the output from the photodetection measurement is in fact essential to know the corresponding conditional state, and thus to extract the whole information on $\omega$ via the final strong measurement.

These results can be readily extended to the scenario of non-linear quantum metrology, where $k$-body Hamiltonian and $p$-body dissipators are considered ~\cite{Beau2017}. If we focus on the case where $k$ and $p$ are odd numbers, by preparing the initial state in a GHZ state, and by monitoring via photodetection each dissipation channel, one can show by a simple rescaling of the variables $\omega$ and $\kappa$ that for unit monitoring efficiency, the ultimate (noiseless) scaling $\sim N^k$ can be restored; on the other hand, as soon as the efficiency is smaller than one, one goes back to the noisy scaling $\sim N^{k-p/2}$, with only a constant enhancement, due to the finite monitoring efficiency.

The results for the case of continuous homodyne monitoring are not reported here since the corresponding effective QFI, at fixed efficiency $\eta$, is always lower than the one obtained for continuous photodetection, and Heisenberg scaling is not recovered even in the case of perfect monitoring.
\subsection{Transverse noise}

Thanks to the symmetry of the GHZ state, the QFI for the unconditional dynamics with arbitrary collapse operators can be obtained without the need to diagonalize the full density matrix~\cite{Chaves2013}.
The corresponding optimized QFI can be then numerically obtained and the scaling is found to be intermediate between SQL and Heisenberg: $\max_t \left[ \qfi \left[ \varrho_\mathsf{unc}^\bot \right] / t \right] \approx N^{5/3} $.

On the other hand, the ultimate QFI can be computed analytically (see Sec.~\ref{s:UQFI} for details on the calculation), yielding
\begin{align}
\label{eq:QFImolmerGHZfinal}
&\overline{\qfi}_{\mathcal{L}_{\omega}}^\bot = \frac{ N^2 \left(1- e^{-\kappa  t}\right)^2 + N \left[ 2 \kappa  t+ 1 - \left(2 - e^{-\kappa t} \right)^2 \right] }{\kappa ^2}.
 \end{align}
Two main observations are in order here: on the one hand we observe that $\overline{\mathcal{Q}}_{\mathcal{L}_{\omega}}^\bot$ depends on the noise parameter $\kappa$ and is always smaller than the noiseless QFI $\mathcal{Q}_{\sf HL}=N^2t^2$.
This shows how, unlike in the parallel noise case, for transverse (non-commuting) noise, part of the information leaking into the environment is irretrievably lost, and cannot be recovered even if one has at disposal all the environmental degrees of freedom.
On the other hand, this expression does explicitly show Heisenberg scaling, and can be further optimized over the evolution time $t$.
In contrast to the unconditional case, the optimal time $t_\mathsf{opt} (N)$ does not go to zero for $N \to \infty $, instead it tends to a constant: $\lim_{N \to \infty} t_\mathsf{opt} (N) = c / \kappa$, where $c \approx 1.26$ .
The very same results can be obtained by computing the unconditional QFI in the limit $\omega \to 0$~\cite{Albarelli2018thesis}, which is already known to give rise to Heisenberg scaling (see Appendix D of~\cite{Brask15}), i.e.
\begin{align}
\overline{\qfi}_{\mathcal{L}_{\omega}}^\bot&= \lim_{\omega \to 0} \qfi \left[ \varrho^\bot_\mathsf{unc} \right].
 \end{align}

\begin{figure}
\includegraphics{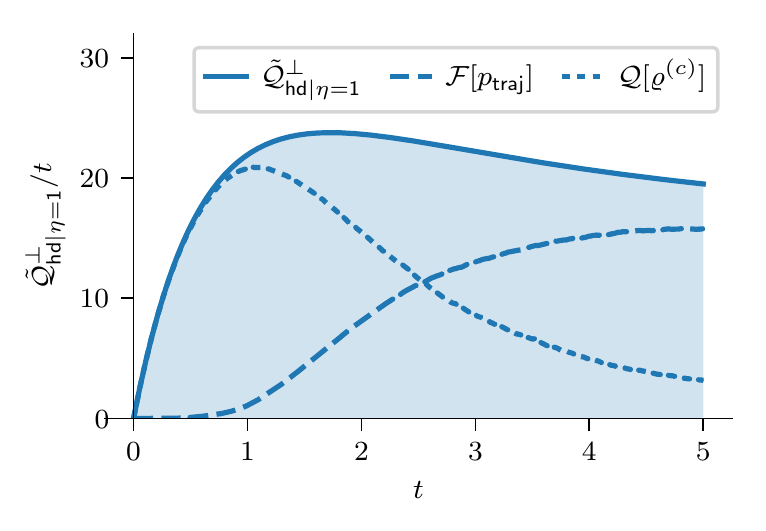}
\caption{Contributions of the classical FI $\mathcal{F}[p_\mathsf{traj}]$ and the QFI of the final strong measurement $\qfi[\varrho^{(c)}]$ to the effective QFI $\tilde{\mathcal{Q}}_{\mathsf{hd|\eta=1}}$ for homodyne with $N=7$ and $\omega = 1$, in the transverse noise case.}
\label{fig:contributions}
\end{figure}

Notice that it is necessary to take the limit, instead of using directly $\omega=0$, because for this value of the parameter the density matrix changes its rank and this gives rise to a discontinuity in the QFI~\cite{Safranek2017,SevesoTBA}.
The effective QFI for photodetection and homodyne detection is obtained numerically with the methods presented in Sec.~\ref{s:algorithm}.
From our results we observe that for unit efficiency $\eta = 1$ the effective QFI saturates the ultimate bound in both cases
\begin{align}
\widetilde{\qfi}^\bot_{\mathsf{hd},\eta=1} = \widetilde{\qfi}^\bot_{\mathsf{pd},\eta=1} =  \overline{\qfi}_{\mathcal{L}_\omega}^\bot.
\end{align}
This has been checked up to $N=14$, but we conjecture that this equality holds in general.
The two terms that contribute to $\widetilde{\qfi}^\bot_{\mathsf{hd},\eta=1}$ in Eq.~\eqref{eq:effQFI} always sum up to $\overline{\qfi}_{\mathcal{L}_\omega}^\bot$ but with $\omega$-dependent behaviors.
This is shown for a particular set of parameters in Fig.~\ref{fig:contributions}.
On the other hand, as explained before for the parallel case, $\widetilde{\qfi}^\bot_{\mathsf{pd},\eta=1}$ is only equal to the average QFI of the conditional states, since the classical FI  $\mathcal{F}[p_\mathsf{traj}]$ is always identical to zero.
As expected, for $\eta < 1 $, we observe that the effective QFI lies between the unconditional QFI and the ultimate QFI, confirming the chained inequalities~\eqref{eq:QFIineq} for both detection strategies and also the conjecture regarding the monotonicity with the efficiency $\eta$.
\begin{figure*}
	\includegraphics{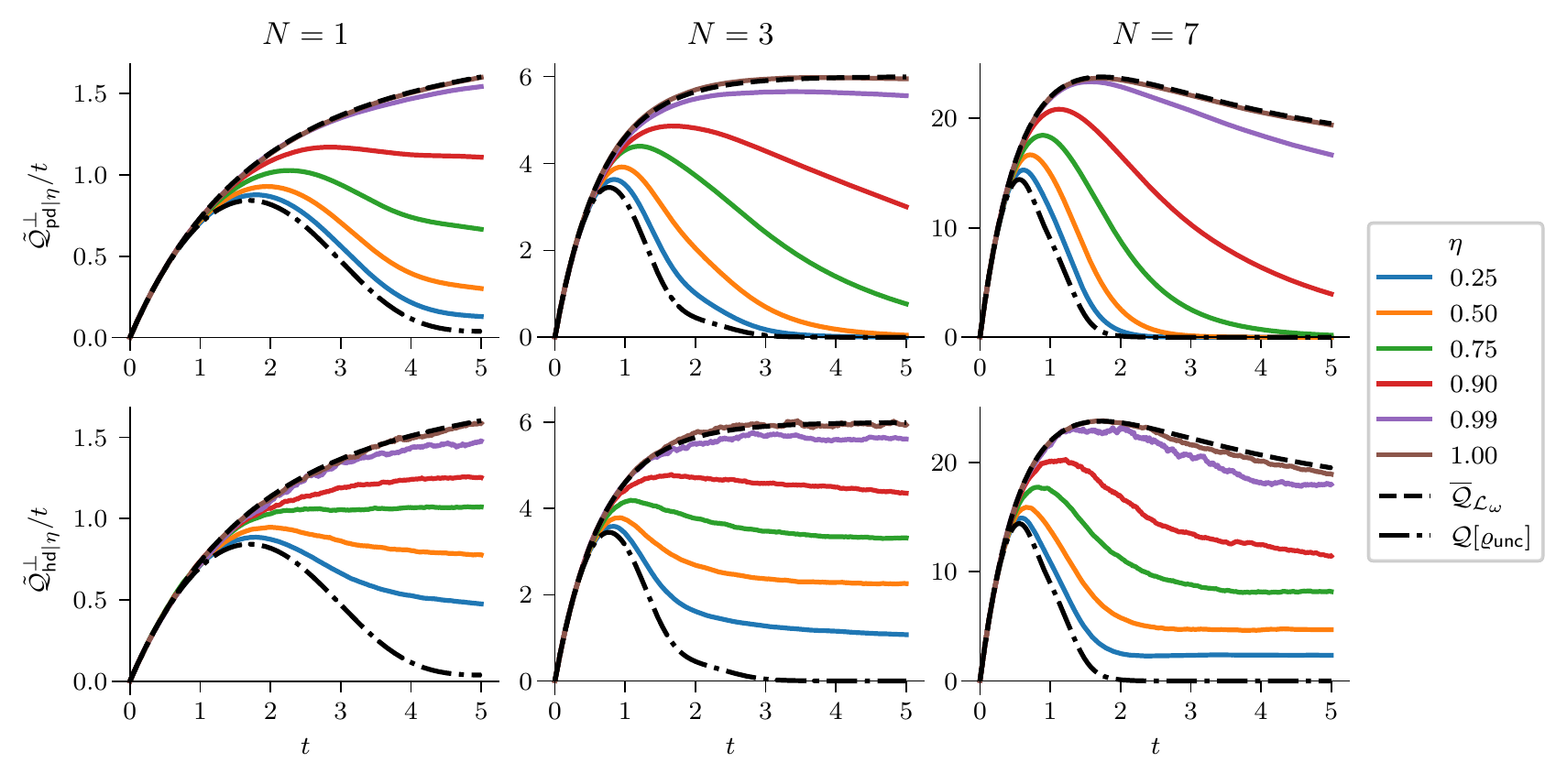}
	\caption{The figure shows the effective QFI for photodetection $\widetilde{\qfi}^\bot_{\mathsf{pd}, \eta}/t$ (top row) and for homodyne  $\widetilde{\qfi}^\bot_{\mathsf{ph}, \eta}/t$ (bottom row) with transverse noise over time, for different efficiencies $\eta$ and for different values of $N$.
	The effective QFI is compared with the ultimate bound $\overline{\qfi}_{\mathcal{L}_\omega}^\bot$ (black dashed line), and the QFI for the unconditional evolution $\qfi[\varrho^\bot_\mathsf{unc}]$ (black dot-dashed line).
Here we take $\omega = \kappa$.	The colored curves are obtained numerically as explained in Sec.~\ref{s:algorithm}, by simulating a large number $> 10 \,k$ trajectories.}
	\label{fig:qeff_over_t_vs_t}
\end{figure*}

As an example, in Fig.~\ref{fig:qeff_over_t_vs_t} we show the effective QFI over time for photodetection and homodyne detection at different efficiencies, for three different values of $N$.
We can see that at lower efficiencies the curves tend to the unconditional QFI $\qfi[\varrho_\mathsf{unc}^\bot]$, while for perfect efficiency they coincide with $\overline\qfi_{\mathcal{L}_\omega}$.
Notice that in general one cannot define a hierarchy between the two strategies, and that in particular, in the case of homodyne detection, the curves become constant at large $t$, due to the non-vanishing contribution of the classical Fisher information $\mathcal{F}[p_\mathsf{traj}]$ that is linear in $t$.

\begin{figure*}
	\includegraphics{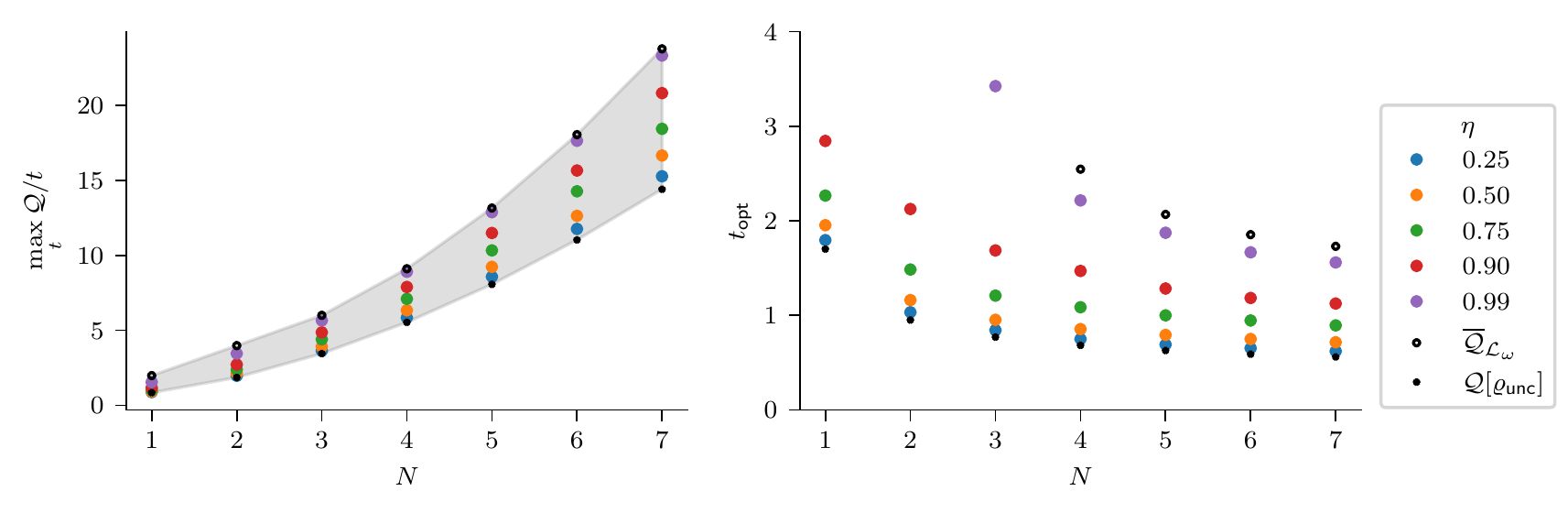}
	\caption{The maximum over time of the quantity $\qfi/t$ (left panel) and the corresponding optimal time (right panel) for $\omega=\kappa$, where $\qfi = \widetilde{\qfi}^\bot_{\mathsf{pd|\eta}}$ for various values of $\eta$ for the colored dots, $\qfi = \overline{\qfi}^\bot_{\mathcal{L}_\omega}$ for the black circles,
	and $\qfi = \qfi[\varrho_\mathsf{unc}]$ for the black stars.
The gray shade shows the region individuated by the bounds in Eq.~\eqref{eq:QFIineq}.
	Notice that $ \overline{\qfi}^\bot_{\mathcal{L}_\omega}/t$ is monotonically increasing for $N \leq 3$, and it reaches its maximum value for $t\to \infty$, and that the optimal measurement time $N \leq 2$ for $\eta = 0.99$ is out of scale.}
	\label{fig:q_opt_pd}
\end{figure*}

Since the numerical method for non-unit efficiency requires using the full Hilbert space,
the complexity of the algorithm is exponential in $N$ and we have been able to obtain
results only up to $N=7$.
Consequently, we cannot explicitly witness a different scaling from the unconditional case.
As a matter of fact the difference between $\max_t \left[ \overline{\qfi}_{\mathcal{L}_\omega}^\bot / t \right]$
and $\max_t \left[ \qfi \left[ \varrho^\bot_\mathsf{unc} \right] /t \right]$ is not very significant
for $N \leq 7$. This is shown in Fig.~\ref{fig:q_opt_pd} for photodetection, in the case $\omega = \kappa$.
As we can see, the two quantities have a similar scaling in this range of $N$, with the effective QFI lying between them with optimal values that are monotonous with $\eta$.
The optimal measurement time decreases with $N$, and it increases with increasing $\eta$.

\section{Evaluation of the ultimate QFI} \label{s:UQFI}
In this section we show how to derive the analytical formulas for the ultimate QFI $\overline{\qfi}_{\mathcal{L}_\omega}$ for both parallel and transverse noise.
\subsection{Parallel noise}
We start by showing that, for parallel noise, the ultimate QFI is equal to the QFI of the noiseless case: the proof is however more general and is valid whenever the collapse operators $\hat{c}_j$ commute with the free Hamiltonian $\hat{H}_\omega$.

The master equation~\eqref{eq:Markov} is obtained by considering the following interaction Hamiltonian between the system and the input modes: $\hat{H}_\text{int} (t) = \sum_{j=1}^N \left( \hat{c_j} \hat{a}^{(j) \dag}_\mathsf{in} (t) + \hat{c}^\dag_j \hat{a}^{(j)}_\mathsf{in} (t) \right)$.
We remark that $t$ is merely a parameter that labels which mode interacts with the system at time $t$ and for each $t$ we have a different operator acting on a different Hilbert space.
The total time-dependent Hamiltonian for the system and environment is thus $\hat{H}_\mathsf{SE} (t) = \hat{H}_\omega + \hat{H}_\mathsf{int} (t) $ and each input mode interacts with the main system for an infinitesimal time $d t$.
However, for the sake of clarity, we will consider a discretization with a finite interaction time $\delta t$, so  that the evolution over a total time $T= M \delta t$ involves a finite number $M$ of input modes.
We also assume that the state of the input modes is the vacuum $|0 \rangle$ and that the initial state of the system $| \psi_0 \rangle$ is pure.

Under these assumptions the joint state of system and environment evolves as
\begin{equation}
|  \psi_\mathsf{SE} (\omega) \rangle = \hat{U}_{t_M} \dots \hat{U}_{t_2} \hat{U}_{t_1} \left( | \psi_0 \rangle \otimes | 0 \rangle^{\otimes M} \right),
\end{equation}
where $t_j = j \cdot \delta t $ and $U_{t_j}= \exp \left[ -i \delta t \left( \hat{H}_{\omega} + \hat{H}_\text{int} (t_j) \right)  \right]$.

This joint state is connected to the operator $\bar{\varrho}$ appearing in Eq.~\eqref{eq:MolmerGenME}, since its trace represents the fidelity for two different values of the parameter~\cite{GammelmarkQCRB,Macieszczak2016}, i.e.
\begin{equation}
\label{eq:overlapQFImolmer}
\langle \psi_\mathsf{SE}  (\omega_1) |  \psi_\mathsf{SE}  (\omega_2) \rangle =\Tr\left[\bar \varrho \right].
\end{equation}

When all the collapse operators $\hat{c}_j$ commute with the free Hamiltonian $\hat{H}_\omega$, $\hat{H}_\mathsf{int}$ commutes  as well and we have
\begin{equation}
\hat{U}_{t_i}= \exp \left[ - i \delta t \hat{H}_\mathsf{int} (t_i) \right] \cdot \exp \left[ -i \delta t \hat{H}_{\omega}  \right].
\end{equation}
Therefore, in the computation of the overlap~\eqref{eq:overlapQFImolmer} the terms due to the interaction cancel out and we have
\begin{equation*}
\langle \psi_\mathsf{SE}  (\omega_1) |  \psi_\mathsf{SE}  (\omega_2) \rangle = \langle \psi_0 | \exp\left[ -i T \left( H_{\omega_2} - H_{\omega_1} \right) \right] | \psi_0 \rangle,
\end{equation*}
so that Eq.~\eqref{eq:ultimQFI} gives the QFI of the unitary case.
In particular, this is true for parallel noise, i.e., for $\hat{c}_j=\sqrt{k/2} \sigma_z^{(j)}$, and, for any pure initial state of the system $|\psi_0\rangle$, we have:
\begin{align}
 \overline{\qfi}_{\mathcal{L}_{\omega}}^{\mathsf{\parallel}} &= \qfi \left[ e^{-i \hat{H}_\omega t} | \psi_0\rangle \!\langle \psi_0 |  e^{i \hat{H}_\omega t}\right]  \\
 &= 4 \left[ \langle \psi_0 | \left( \partial_\omega \hat{H}_\omega \right)^2 | \psi_0\rangle - \langle \psi_0 | \left( \partial_\omega \hat{H}_\omega \right) | \psi_0\rangle^2 \right]\,. \notag
\end{align}

\subsection{Transverse noise}
When the collapse operators do not commute with the generator, as in the case of frequency estimation with transverse noise, we can simplify the computation by using the assumption of an identical and independent noise acting on each qubit.
Since Eq.~\eqref{eq:MolmerGenME} is linear in $\bar{\varrho}$ and the coefficients are time-independent, we can still write the solution as a linear map, formally $\mathcal{\tilde{E}}_{\omega_1, \omega_2}(t) = \exp \left( t \mathcal{\tilde{L}}_{\omega_1,\omega_2}  \right) $.
This map is only guaranteed to be linear and in general it is not even positive.

Since the map acts independently on every qubit, we can still write the global action on the $N$-qubit state as the tensor product $\mathcal{\tilde{E}}^N_{\omega_1,\omega_2} (t) = \mathcal{\tilde{E}}_{\omega_2,\omega_2}(t)^{\otimes N}$, where $\mathcal{\tilde{E}}_{\omega_2,\omega_2}(t)$ is the single-qubit solution.
The ultimate bound is thus obtained as
\begin{equation}
\label{eq:MolmerFromChannels}
\overline{\qfi}_{\mathcal{L}_{\omega}} \left[ \ket{\psi_0} \right] = 4 \left. \partial_{\omega_1} \partial_{\omega_2} \log \Tr [ \mathcal{\tilde{E}}^N_{\omega_1,\omega_2} (t) \, \varrho_0 ] \right|_{\omega_1 = \omega_2 = \omega},
\end{equation}
where $\varrho_0$ is the initial pure state.
Given our choice $\varrho_0 = |\psi_\mathsf{GHZ}\rangle\!\langle \psi_\mathsf{GHZ}| $, the computation can be greatly simplified.
We find that
\begin{align}
&\mathcal{\tilde{E}}^N_{\omega_1,\omega_2} (t) \,  |\psi_\mathsf{GHZ}\rangle\!\langle \psi_\mathsf{GHZ}|= \\ \notag
&= \frac{1}{2} \Biggl[ \left( \mathcal{\tilde{E}}_{\omega_1,\omega_2} (t) \, |0 \rangle \! \langle 0 | \right)^{\otimes N} + \left( \mathcal{\tilde{E}}_{\omega_1,\omega_2} (t) \, |1 \rangle \! \langle 1 | \right)^{\otimes N} \\ \notag
& \quad \; \, + \left( \mathcal{\tilde{E}}_{\omega_1,\omega_2} (t) \, |0 \rangle \! \langle 1 | \right)^{\otimes N} + \left( \mathcal{\tilde{E}}_{\omega_1,\omega_2} (t) \, |1 \rangle \! \langle 0 | \right)^{\otimes N}   \Biggr].
\end{align}

For transverse noise and for a single qubit, the equation to solve to compute $\overline{\qfi}$ is the following
\begin{equation}
\begin{split}
\frac{d \bar{\varrho}}{d t} &= \mathcal{ \tilde{L} }^{\bot}_{\omega_1,\omega_2} \left[ \bar{\varrho} \right]  =\\
 & = -\frac{i}{2} \left( \omega_1  \sigma_z \bar{\varrho}  - \omega_2 \tilde{\varrho} \sigma_z  \right) + \frac{\kappa}{2} \left( \sigma_x \bar{\varrho} \sigma_x - \bar{\varrho} \right).
 \end{split}
\end{equation}
The solution is obtained in the same way as for canonical master equations: by choosing a basis of operators and using a matrix representation of superoperators~\cite{Andersson2007} (see also the Supplementary Material of~\cite{GammelmarkQCRB}).
By making use of the normalized Pauli operators $\tilde{\sigma}_i = \sigma_i / \sqrt{2}$ (where $\sigma_0 = \id$), so that $\Tr \left[ \tilde{\sigma}_i \tilde{\sigma}_j \right] = \delta_{ij} $, we can find the matrix associated to the single qubit map $\mathcal{\tilde{E}}_{\omega_1,\omega_2} (t)$, obtained by matrix exponentiation.
We can write the generalized density operator $\bar{\varrho}$ in Bloch form as $\bar{\varrho}= \frac{1}{\sqrt{2}} \left( a_0 \tilde{\sigma}_0 + \vec{a} \cdot \vec{\tilde{\sigma}} \right) $ such that its trace is simply $\Tr \left[ \bar{\varrho} \right] = a_0$.
In this notation, the initial states $| 0 \rangle \! \langle 0 |$ and $ | 1 \rangle \! \langle 1 |$
correspond to the vectors $e_{00} = \frac{1}{\sqrt{2}} (1,0,0,1)^\mathsf{T}$ and $e_{11} = \frac{1}{\sqrt{2}} (1,0,0,-1)^\mathsf{T}$, while the off-diagonal elements are \(e_{01/10}= \frac{1}{\sqrt{2}} (0,1,\pm i,0)^\mathsf{T} \).

Since we are only interested in the coefficient $a_0$, we only need the first row of the matrix representation of $\mathcal{\tilde{E}}^{\bot}_{\omega_1,\omega_2}(t)$, which turns out to be

\begin{widetext}
\begin{equation}
	\resizebox{.9\textwidth}{!}{$
	\left[ \mathcal{\tilde{E}}^{\bot}_{\omega_2,\omega_2}(t) \right]_{0,*}=  e^{-\frac{\kappa t}{2}} \left(  \cosh \left( \frac{t}{2} \sqrt{ \kappa^2 - \left(\omega _1-\omega _2\right)^2 }\right)
	+
	 \frac{\kappa  \sinh \left( \frac{t}{2} \sqrt{\kappa^2 - \left(\omega _1-\omega _2\right)^2 }\right)}{2 \sqrt{ \kappa^2 - \left(\omega _1-\omega _2\right)^2 } } , 0 , 0 ,
	 \frac{ i	 \left(\omega _2-\omega _1\right)  \sinh \left(\frac{t}{2} \sqrt{\kappa^2 - \left(\omega _1-\omega _2\right)^2 } \right) }{\sqrt{\kappa^2 - \left(\omega _1-\omega _2\right)^2 } } \right)$} .
\end{equation}
\end{widetext}
We can see that the off-diagonal terms are kept off-diagonal by the map, therefore they do not contribute to the trace and we can further simplify the calculation as
\begin{align}
\label{eq:QFImolmerGHZpre}
 &\Tr \left[ \mathcal{\tilde{E}}^{\bot N}_{\omega_1,\omega_2} (t)  |\psi_\mathsf{GHZ}\rangle\!\langle \psi_\mathsf{GHZ}|  \right] = \\
 & = \frac{1}{2} \left( \Tr \left[ \mathcal{\tilde{E}}^{\bot}_{\omega_1,\omega_2} (t) |0 \rangle \! \langle 0 | \right]^{ N}
 + \Tr \left[ \mathcal{\tilde{E}}^{\bot}_{\omega_1,\omega_2} (t) |1 \rangle \! \langle 1 | \right]^{N} \right) \notag \\
 & = \frac{1}{\sqrt{2}} \Biggl\{
  \left( [\mathcal{\tilde{E}}^{\bot}_{\omega_1,\omega_2}(t)]_{0,*} \cdot e_{00} \right)^N  + \left( [ \mathcal{\tilde{E}}^{\bot}_{\omega_1,\omega_2}(t)]_{0,*} \cdot e_{11} \right)^N   \Biggr\}. \notag
\end{align}
We can thus plug this result into~\eqref{eq:MolmerFromChannels} and finally obtain Eq.~\eqref{eq:QFImolmerGHZfinal}.
\section{Numerical algorithm for the calculation of the effective QFI for time-continuous strategies} \label{s:algorithm}
We can now describe the numerical algorithm we have implemented to calculate the effective QFI $\widetilde{\qfi}_\mathsf{unr,\eta}$.
The numerical results presented in this manuscript are obtained with the code available online at~\cite{ContinuousMeasurementFI}, written in the Julia language~\cite{julia}.

We will focus on the case of homodyne detection and then we will discuss the small changes needed for photodetection.
We first review the existing method to calculate one of the two key quantities in Eq.~\eqref{eq:effQFI}, namely the classical Fisher information $\mathcal{F}[p_\mathsf{traj}]$.
In \cite{GammelmarkCRB} it is shown that it can be calculated as
\begin{align}
\mathcal{F}[p_\mathsf{traj}] = \mathbbm{E}\left[ \Tr[\tau]^2 \right] , \label{eq:Fishtau}
\end{align}
where we have defined the operator
 \begin{align}
\tau= \frac{ \partial_\omega \tilde{\varrho}^{(c)} }{\Tr[\tilde{\varrho}^{(c)}]} \:, \label{eq:tau_def}
\end{align}
 in terms of the unormalized conditional state $\tilde{\varrho}^{(c)}$ evolving according to the SME
 \begin{align}
 d\tilde{\varrho}^{(c)} = {}& - i [\hat{H}_{\omega} , \tilde{\varrho}^{(c)}]\,dt + \sum_j \mathcal{D}[\hat{c}_j] \tilde{\varrho}^{(c)} \,dt \notag \\
 & + \sum_j   \sqrt{\eta}_j \left( \hat{c}_j \tilde{\varrho}^{(c)} + \tilde{\varrho}^{(c)} \hat{c}_j^\dagger \right) dy_j \,,  \label{eq:unormalizedSME}
\end{align}
and such that $\varrho^{(c)}= \tilde{\varrho}^{(c)}/\Tr[\tilde{\varrho}^{(c)}]$.

One can then numerically integrate simultaneously the two SMEs, the one for the conditional state $\varrho^{(c)}$ in Eq.~\eqref{eq:homoSME}, and the one above for $\tau$, and then evaluate the corresponding FI, by averaging over a certain number of trajectories.
However, it is computationally very expensive to obtain a guaranteed Hermitian $\varrho^{(c)}$ by using standard methods, such as Euler-Maruyama or Euler-Milstein, and still these methods do not guarantee the positivity of corresponding density operator.

An alternative method to integrate the SME, which is able to circumvent these problems, has been introduced in \cite{Rouchon2014, Rouchon2015}, by exploiting the Kraus operators corresponding to the (weak) measurement performed at each instant of time.
After an infinitesimal time $dt$, the evolved state corresponding to~\eqref{eq:homoSME} can in fact be written as
\begin{align}\label{eq:rouchon}
\varrho^{(c)}_{t+dt} = \frac{ M_\mathbf{dy}^\phd\varrho^{(c)}_t M_\mathbf{dy}^\dag + \sum_j (1-\eta_j) \hat{c}^\phd_j \varrho_t^{(c)} \hat{c}_j^\dag \,dt} { \Tr[ M_\mathbf{dy}^\phd\varrho^{(c)}_t M_\mathbf{dy}^\dag +\sum_j  (1-\eta_j)  \hat{c}^\phd_j \varrho_t^{(c)} \hat{c}_j^\dag \,dt ]} \, ,
\end{align}
where we have explicitly put the time dependence of the density operators and where we have defined the Kraus operator
\begin{align}
M_\mathbf{dy}^\phd= \mathbbm{1} - i \hat{H}_\omega \,dt - \frac{1}{2} \sum_j \hat{c}_j^\dag \hat{c}^\phd_j \, dt + \sum_j \sqrt{\eta} \hat{c}_j \,dy_j \,, \label{eq:Mdyvec}
\end{align}
with $\mathbf{dy} = \{ dy_j \}$ being a vector of measurement results, corresponding to each output channel,
\begin{equation}
	dy_j = \sqrt{\eta_j}\,\Tr[\varrho_t^{(c)} (\hat{c}^\phd_j + \hat{c}_j^\dag)] \,dt + dw_j.
\end{equation}

Equation \eqref{eq:rouchon} can be used for numerical purposes, where the infinitesimal time $dt$ is replaced by a finite time step $\Delta t$, the Wiener increments $dw_j$ are replaced by Gaussian random variables $\Delta w_j$ centered in zero and with variance equal to $\Delta t$, and where one can also implement the second order Euler-Milstein corrections.
The (numerical) Kraus operators in this case read
\begin{align}
M_\mathbf{\Delta y} ={}& \mathbbm{1} - i \hat{H}_{\omega} \, \Delta t - \frac{1}{2}  \sum_j \hat{c}_j^\dag \hat{c}^\phd_j \, \Delta t + \sum_j \sqrt{\eta_j}\, \hat{c}_j \, \Delta y_j  \notag \\
& +  \sum_{j,k} \frac{\eta}{2} \hat{c}_j \hat{c}_k ( \Delta y_j \Delta y_k  - \delta_{j,k} \Delta t) \, ,
\label{eq:numericalKraus}
\end{align}
with \begin{align}
\Delta y_j =  \sqrt{\eta_j}\, \Tr[ \varrho_t^{(c)} (\hat{c}^\dag_j + \hat{c}_j^\phd)] \, \Delta t + \Delta w_j \,,
\end{align}
denoting the (finite) increments of the measurement records.

In the following we will show how to extend this method to obtain an efficient and numerically stable calculation of both $\mathcal{F}[p_\mathsf{traj}]$ and $\qfi[\varrho_c]$.
We will prove everything in terms of the ``infinitesimal'' Kraus operators, which will have to be replaced by Eq.~\eqref{eq:numericalKraus} when implementing the numerical algorithm.
We will start by showing the results for the most general case of SME and inefficient detection; we will then describe the more efficient algorithm that can be implemented in the case of perfect detection ($\eta_j=1$), i.e., when the dynamics can be described by a stochastic Schr\"odinger equation.

\subsection{Non-unit efficiency detection  (stochastic master equation)}
We start by observing that the evolution of the unnormalized conditional state and of its derivative can be written in terms of the Kraus operators
(in what follows we will omit the superscript $(c)$ used to denote conditional states):
\begin{align}
\tilde{\varrho}_{t+dt} ={}& M_\mathbf{dy}^\phd\tilde{\varrho}_t M_\mathbf{dy}^\dag + \sum_j (1-\eta_j) \hat{c}_j^\phd \tilde{\varrho}_t \hat{c}_j^\dag \,dt  \:, \notag \\
\partial_\omega \tilde{\varrho}_{t+dt} ={}& M_\mathbf{dy}^\phd(\partial_\omega \tilde{\varrho}_t) M_\mathbf{dy}^\dag
+ (\partial_\omega M_\mathbf{dy}^\phd) \tilde{\varrho}_t M_\mathbf{dy}^\dag + \notag \\
& + M_\mathbf{dy}^\phd\tilde{\varrho}_t (\partial_\omega M_\mathbf{dy}^\dag )  +\notag \\
& + \sum_j (1-\eta_j) \hat{c}_j^\phd (\partial_\omega \tilde{\varrho}_t) \hat{c}_j^\dag \,dt\:, \label{eq:rho_tilde_kraus}
\end{align}
where $\partial_\omega M_\mathbf{dy}^\phd= -i (\partial_\omega \hat{H}_\omega) \,dt$.
The trace of the unnormalized state reads
\begin{align}
\Tr[\tilde{\varrho}_{t+dt} ] &= \Tr[M_\mathbf{dy}^\phd\tilde{\varrho}_t  M_\mathbf{dy}^\dag + \sum_j (1-\eta_j) \hat{c}_j^\phd \tilde{\varrho}_t  \hat{c}_j^\dag \,dt ]  \notag \\
&= \Tr[\tilde{\varrho}_t] \Tr[M_\mathbf{dy}^\phd\varrho_t  M_\mathbf{dy}^\dag + \sum_j (1-\eta_j) \hat{c}_j^\phd {\varrho}_t  \hat{c}_j^\dag \,dt ]\,, 
\end{align}
where we have used the relation $\varrho_t = \tilde{\varrho}_t / \Tr[\tilde{\varrho}_t]$.
We can now use these formulas to obtain
the evolution for the operator $\tau_{t+dt}$ just in terms of the operators $\varrho_t$ and $\tau_t$ at the previous time step:
\begin{widetext}
\begin{align}
\tau_{t+dt}
={} & \frac{1}{\Tr[\tilde{\varrho}_{t+dt}]} \left[(\partial_\omega M_\mathbf{dy}^\phd ) \tilde{\varrho}_t M_\mathbf{dy}^\dag +
 M_\mathbf{dy}^\phd(\partial_\omega \tilde{\varrho}_t) M_\mathbf{dy}^\dag + M_\mathbf{dy}^\phd \tilde{\varrho}_t (\partial_\omega M_\mathbf{dy}^\dag) + \sum_j (1-\eta_j) \hat{c}^\phd_j (\partial_\omega \tilde{\varrho}_t ) \hat{c}_j^\dag    \,dt \right] \notag  \\
={} & \frac{(\partial_\omega M_\mathbf{dy}^\phd ) \varrho_t M_\mathbf{dy}^\dag + M_\mathbf{dy}^\phd\tau_t M_\mathbf{dy}^\dag +
M_\mathbf{dy}^\phd\varrho_t (\partial_\omega M_\mathbf{dy}^\dag) + \sum_j (1-\eta_j) \hat{c}^\phd_j \tau_t \hat{c}_j^\dag   \,dt}
{\Tr\left[M_\mathbf{dy}^\phd\varrho_t  M_\mathbf{dy}^\dag + \sum_j (1-\eta_j) \hat{c}^\phd_j {\varrho}_t \hat{c}_j^\dag \,dt \right]} \,.
\label{eq:tau}
\end{align}
\end{widetext}

One can thus evaluate the trace of this operator at each time $t$, and evaluate accordingly the classical Fisher information $\mathcal{F}[p_\mathsf{traj}]$ as in Eq.~\eqref{eq:Fishtau}.

Notice that the evolution for the derivative operator $\partial_\omega \varrho_t$ can be now written in terms of the renormalized operators $\varrho_t$ and $\tau_t$:
\begin{align}
\partial_\omega \varrho_t &= \frac{\partial_\omega \tilde{\varrho}_t }{\Tr[\tilde{\varrho}_t]} -
\frac{\Tr[\partial_\omega \tilde{\varrho}_t]}{\Tr[\tilde{\varrho}_t]^2} \tilde{\varrho}_t = \tau_t - \Tr[\tau_t] \varrho_t \,. \label{eq:drho}
\end{align}

The QFI $\qfi[\varrho_t]$ can then be evaluated at each time $t$ by using the formula \cite{MatteoIJQI}:
\begin{align}
\qfi[\varrho_t] = 2\sum_{\lambda_s + \lambda_t \neq 0} \frac{|\langle \psi_s |
\partial_\omega \varrho_t | \psi_t \rangle |^2}{\lambda_s + \lambda_t} \,,
\end{align}
upon writing $\varrho_t$ in its eigenbasis, i.e., $\varrho_t =\sum_s \lambda_s |\psi_s\rangle\!\langle\psi_s| \,.$

We would like to underline the key features of our algorithm.
The relevant figures of merit could naively be derived from the evolution of the unnormalized conditional state
 $\tilde\varrho_t$, described in Eq.~\eqref{eq:rho_tilde_kraus}.
However $\Tr[\tilde\varrho_t]$ becomes very small during the evolution, leading to numerical instabilities in the evaluation of both  $\mathcal{F}[p_\mathsf{traj}]$ and $\qfi[\varrho_t]$.
Thanks to Eqs. \eqref{eq:tau} and \eqref{eq:drho}, we are able to express the above quantities only in terms of the numerically stable operators $\varrho_t$ and $\tau_t$.
Besides, the formulation in terms of Kraus operators, following \cite{Rouchon2014,Rouchon2015}, ensures that the density operator remains positive, as opposed to standard numerical integration of the SME.

\subsection{Unit efficiency detection (stochastic Schr\"odinger equation)}
The above calculations are greatly simplified when the dynamics starts with a pure initial state and the efficiency parameters are equal to one, $\eta_j=1$.
In fact the quantum conditional state $|\psi_t\rangle$ remains pure during the whole evolution and the dynamics is described by a stochastic Schr\"odinger equation.
We can thus work with state vectors, instead of density matrices, with a consequent reduction of complexity of the numerical simulation, which allows us to reach higher values of $N$ with a given amount of memory.

In terms of Kraus operators, the unnormalized and normalized conditional states are obtained respectively as:
\begin{align}
|\widetilde\psi_{t+dt} \rangle &= M_\mathbf{dy}^\phd|\widetilde\psi_t \rangle \,, \\
|\psi_{t+dt} \rangle &= \frac{ M_\mathbf{dy}^\phd|\psi_t \rangle }{\sqrt{\langle \psi_t | M_\mathbf{dy}^\dag M^{\phd}_\mathbf{dy} | \psi_t\rangle }} =
\frac{ M_\mathbf{dy}^\phd|\widetilde\psi_t \rangle }{\sqrt{\langle \widetilde\psi_t | M_\mathbf{dy}^\dag M^{\phd}_\mathbf{dy} | \widetilde\psi_t\rangle }}  \notag \,.
\end{align}
The operator $\tau_t$ in this case can be written as
\begin{align}
\tau_t &= \frac{\partial_\omega (  |\widetilde\psi_t \rangle \! \langle  \widetilde\psi_t | )}{
\langle \widetilde\psi_t  |\widetilde\psi_t \rangle}
= \frac{| \partial_\omega \widetilde\psi_t \rangle \! \langle  \widetilde\psi_t |  +| \widetilde\psi_t \rangle \! \langle  \partial_\omega \widetilde\psi_t | }{
\langle \widetilde\psi_t |\widetilde\psi_t \rangle} \,,
\end{align}
 and its trace is equal to
 \begin{align}
 \Tr[\tau_t] &= \frac{ \langle \widetilde\psi_t | \partial_\omega \widetilde\psi_t\rangle + \hbox{h.c.} } {\langle \widetilde\psi_t  | \widetilde\psi_t \rangle}
= \langle \psi_t  | \phi_t\rangle + \langle \phi_t | \psi_t  \rangle \,. \label{eq:tracephi}
 \end{align}
In the last equation we have introduced the vector
\begin{align}
|\phi_t \rangle = \frac{ |\partial_\omega \widetilde \psi_t  \rangle} { \sqrt{\langle \widetilde\psi_t  | \widetilde\psi_t  \rangle}} \,.
\end{align}

At time $t+dt$, the vector $|\phi_{t+dt}\rangle$ can be obtained as
\begin{align}
|\phi_{t+dt} \rangle &= \frac{ (\partial_\omega M_\mathbf{dy}^\phd) |\widetilde\psi_t \rangle  + M_\mathbf{dy}^\phd|\partial_\omega \widetilde\psi_t \rangle }{\sqrt{\langle \widetilde\psi_t | M_\mathbf{dy}^\dag M^{\phd}_\mathbf{dy} | \widetilde\psi_t \rangle}}  \notag \\
&= \frac{ (\partial_\omega M_\mathbf{dy}^\phd) |\psi_t \rangle  + M_\mathbf{dy}^\phd|\phi_t\rangle }{\sqrt{\langle \psi_t  | M_\mathbf{dy}^\dag M^{\phd}_\mathbf{dy}| \psi_t \rangle}}\,, \label{eq:phit}
\end{align}
where we have exploited the identity
\begin{equation}
\langle \widetilde \psi_t  | M_\mathbf{dy}^\dag M^{\phd}_\mathbf{dy} | \widetilde \psi_t \rangle = \langle  \widetilde\psi_t  | \widetilde\psi_t \rangle \! \langle \psi_t |M_\mathbf{dy}^\dag M^{\phd}_\mathbf{dy} | \psi_t \rangle \,.
\end{equation}

We notice that as in Eq.~\eqref{eq:tau}, the evolution equation for the vector $|\phi_{t+dt}\rangle$ depends only on the vectors $|\psi_t \rangle$ and $|\phi_t\rangle$ and not on the unnormalized state $|\widetilde\psi_t \rangle$, and one can readily evaluate the classical Fisher information in terms of these two vectors via Eqs.~\eqref{eq:tracephi} and~\eqref{eq:Fishtau}.

Regarding the QFI of the conditional state, we first observe that
\begin{align}
|\partial_\omega \psi_t\rangle &= \frac{ |\partial_\omega \widetilde{\psi}_t\rangle}{\sqrt{\langle \widetilde\psi_t | \widetilde\psi_t \rangle }} -
\frac{\langle \widetilde \psi_t |\partial_\omega \widetilde\psi_t\rangle +
\langle \partial_\omega \widetilde\psi_t | \widetilde\psi_t\rangle |\widetilde \psi _t\rangle}{2 \langle \widetilde \psi_t | \widetilde\psi_t\rangle^{3/2}} \notag \\
&= |\phi_t\rangle - \frac{\langle \psi_t | \phi_t\rangle + \langle \phi_t | \psi_t\rangle}{2} |\psi_t\rangle \,.
\end{align}
As we are dealing with pure states, the QFI of the conditional state can be simply evaluated as \cite{MatteoIJQI}
\begin{align}
\qfi[|\psi_t\rangle ] = 4 \left[ \langle \partial_\omega \psi_t | \partial_\omega \psi_t\rangle + (\langle \partial_\omega \psi_t | \psi_t\rangle)^2 \right] \,.
\end{align}

The algorithms above have been derived for the case of time-continuous homodyne detection.
Nonetheless one can easily extend them to time-continuous photodetection as follows.

The Kraus operators $M_\mathbf{dy}^\phd$ have to be replaced at each time step with one of the following Kraus operators corresponding to ``no detector click'' and ``detector click'' (in one of the output channels) events, respectively
\begin{align}
\label{eq:M0pd} M_0 &= \mathbbm{1} - i \hat{H}_\omega \, dt - \frac{1}{2} \sum_j \hat{c}_j^{\dag} \hat{c}^{\phd}_j \,dt \, ,\\
\label{eq:M1pd} M_1^{(j)} &= \sqrt{\eta_j\, dt} \, \hat{c}_j \,.
\end{align}
At each time step, each Kraus operator $M_1^{(j)}$ has to be applied with probabilities $p_1^{(j)} = \eta_j \Tr[\varrho_t \hat{c}_j^\dag \hat{c}^{\phd}_j ]\,dt$ and, correspondingly, the Kraus operator $M_0$ has to be applied with probability $p_0 = 1 - \sum_j p_1^{(j)}$ \cite{WisemanMilburn}.

We finally remark that in the case of frequency estimation with parallel noise and initial GHZ state, the numerical calculations are incredibly simplified thanks to the symmetry of the dynamics.
In fact the whole (both conditional and unconditional) evolution is equivalently described in two two-dimensional Hilbert space spanned by the vectors $|\bar 0 \rangle = |0\rangle^{\otimes N}$ and $|\bar 1 \rangle = |1 \rangle^{\otimes N}$.
As regards continuous photodetection, the Kraus operators \eqref{eq:M0pd} and \eqref{eq:M1pd} are replaced by the effective Kraus operators
\begin{align}
\overline{M}_0 &= \mathbbm{1}_2 - i N \omega \overline{\sigma}_z \, dt - (N \kappa/4) \mathbbm{1}_2 \,dt \,,\\
\overline{M}_1 &= \sqrt{(\eta \kappa/2) \, dt} \; N \overline{\sigma}_z \,.
\end{align}
where $\mathbbm{1}_2$ and $\overline{\sigma}_z$ are respectively the identity operator and the Pauli-z matrix in the Hilbert space spanned by $|\bar{0}\rangle$ and $|\bar{1}\rangle$, and where the operators $\overline{M}_1$ and $\overline{M}_0$ have to be applied respectively with probabilities $\bar{p}_1 = N (\eta \kappa/2) \, dt$ (i.e., the total probability of observing a photon in one of the $N$ output channels) and $\bar{p}_0 = 1 - \bar{p}_1$.
More practically the results for generic $N$ qubits can be readily obtained by exploiting the results obtained for one qubit, and rescaling the time as $t \rightarrow Nt$.
In this case it is also easy to show that the efficiency parameter $\eta$ corresponds to the product between the factual efficiency of the photo-detectors (that we assume to be equal) and the fraction of qubits that are effectively monitored.

\section{Conclusions and remarks} \label{s:conclusion}

We have discussed for the first time the usefulness of time-continuous monitoring in the context of noisy quantum metrology.
We have proven that the desired Heisenberg scaling can in fact be restored by exploiting these schemes and we have obtained several conceptually and practically relevant results.\\
We have shown the fundamental difference between parallel and transverse noise with regards to the ultimate limit achievable when the environmental degrees of freedom can be measured.
In the first case, i.e., when the Hamiltonian and the noise generator commute, having access to every degree of freedom of the environment allows us to restore the unitary (noiseless) Heisenberg limit.
In the latter case, i.e., in the presence of transverse noise, the ultimate QFI still presents a Heisenberg (quadratic) scaling in the number of qubits $N$;
however, it is always lower than the unitary QFI, thus showing that some information is irremediably lost due to the interaction with the environment.
These findings complement the results obtained for frequency estimation with open quantum systems~\cite{Haase2018}, where the geometry of the noise is indeed crucial in determining the different achievable scalings. 

The second non-trivial and conceptually relevant result is that estimation strategies based on (sequential) time-continuous monitoring and final strong measurements on the system are optimal, i.e., the corresponding effective QFIs $\qunr{\eta_j}$ are equal to the ultimate QFIs $\overline{\qfi}_\mathcal{L_\omega}$.
This is true for both parallel and transverse noise, showing that no complicated estimation strategies based on ``entangled measurements'' on the environment and system are needed to achieve the ultimate limit on the precision.
This result, which was not suggested by any mathematical or physical intuition, is particularly important from a practical point of view, as it greatly relaxes the assumptions and demands that are requested to achieve the ultimate limits.

We have also discussed in detail the case of finite efficiency monitoring.
In the presence of parallel noise we have shown that the estimation precision is subject to the standard quantum limit, with a behaviour ruled by a reduced effective dephasing: one can still obtain a constant enhancement, which can be made arbitrarily high by increasing the monitoring efficiency. In the presence of transverse noise, we have observed the expected enhancement with respect to the optimized (super-classical) unconditional results \cite{Chaves2013}. However, due to the computational complexity of the algorithm needed to obtain the effective QFI, we could not infer any conclusion regarding the scaling with the number of probes $N$.

It is important to remark that the use of continuous measurements and feedback techniques has long been recognized as a useful tool for fighting decoherence and to prepare non-classical quantum states~\cite{WisemanMilburn,Ahn2003,Ahn2004,Akerman2012,Ganesan2007,Szigeti2014,Tempura,Levante}.
Our metrological scheme follows this line of thought, but with the great advantage that it is based on continuous monitoring only, without error correction steps (or feedback), differing it from other recent approaches in noisy quantum metrology~\cite{Plenio2016,Gefen2016}.

Moreover, while most of the literature on parameter estimation with continuous measurements focuses on the information gained from the continuous signal, the crucial part that makes our protocol able to recover Heisenberg scaling is the final strong measurement on the conditional state.
A great deal of effort has been also devoted to studying the asymptotic properties of estimation via repeated/continuous measurements~\cite{Burgarth2015a,Catana2015,Guta2016}.
In this approach one is usually interested in performing a single run of the experiment and thus observes the system for a long time.
Our point of view is radically different and rooted in previous works on quantum frequency estimation.
As a matter of fact, we are interested in the initial part of the dynamics, long before reaching the steady state.
For this reason the asymptotic regime of the statistical model is reached by running the experiment many times, instead of observing the system for a long time.

In order to obtain these results we have developed a stable algorithm for the evaluation of the effective quantum Fisher information that quantifies the performance of the proposed strategies.
The flexibility of this algorithm and the originality of our protocols pave the way for a series of future investigations that will exploit techniques and ideas that so far have only been applied to standard noisy metrology schemes.
For example, one can investigate the performances of these protocols with optimized~\cite{Frwis2014} and noisy quantum probes~\cite{Gorecka2017}, or by considering entangled ancillary systems~\cite{Demkowicz-Dobrzanski2014,Huang2018,Sbroscia2017}.
Furthermore, one can also study in detail the role played by quantum coherence~\cite{Giorda2016} and the possibility of considering unravellings of non-Markovian master equations~\cite{Diosi2014}.

\begin{acknowledgments}
The authors acknowledge useful discussions with A. Del Campo, M. Paris, P. Rouchon, A. Smirne and T. Tufarelli. MACR was supported by the Horizon 2020 EU collaborative project QuProCS (Grant Agreement No. 641277) and by the Academy of Finland Centre of Excellence program (project 312058). MGG acknowledges support from Marie Skodowska-Curie Action H2020-MSCA-IF-2015 (project ConAQuMe, grant no. 701154) and from a Rita Levi-Montalcini fellowship of MIUR. \\

\noindent \emph{FA and MACR have equally contributed to this work.}
\end{acknowledgments}

\nocite{apsrev41Control}
\bibliographystyle{apsrev4-1}
\bibliography{library}

\begin{thebibliography}{84}%
\makeatletter
\providecommand \@ifxundefined [1]{%
 \@ifx{#1\undefined}
}%
\providecommand \@ifnum [1]{%
 \ifnum #1\expandafter \@firstoftwo
 \else \expandafter \@secondoftwo
 \fi
}%
\providecommand \@ifx [1]{%
 \ifx #1\expandafter \@firstoftwo
 \else \expandafter \@secondoftwo
 \fi
}%
\providecommand \natexlab [1]{#1}%
\providecommand \enquote  [1]{``#1''}%
\providecommand \bibnamefont  [1]{#1}%
\providecommand \bibfnamefont [1]{#1}%
\providecommand \citenamefont [1]{#1}%
\providecommand \href@noop [0]{\@secondoftwo}%
\providecommand \href [0]{\begingroup \@sanitize@url \@href}%
\providecommand \@href[1]{\@@startlink{#1}\@@href}%
\providecommand \@@href[1]{\endgroup#1\@@endlink}%
\providecommand \@sanitize@url [0]{\catcode `\\12\catcode `\$12\catcode
  `\&12\catcode `\#12\catcode `\^12\catcode `\_12\catcode `\%12\relax}%
\providecommand \@@startlink[1]{}%
\providecommand \@@endlink[0]{}%
\providecommand \url  [0]{\begingroup\@sanitize@url \@url }%
\providecommand \@url [1]{\endgroup\@href {#1}{\urlprefix }}%
\providecommand \urlprefix  [0]{URL }%
\providecommand \Eprint [0]{\href }%
\providecommand \doibase [0]{http://dx.doi.org/}%
\providecommand \selectlanguage [0]{\@gobble}%
\providecommand \bibinfo  [0]{\@secondoftwo}%
\providecommand \bibfield  [0]{\@secondoftwo}%
\providecommand \translation [1]{[#1]}%
\providecommand \BibitemOpen [0]{}%
\providecommand \bibitemStop [0]{}%
\providecommand \bibitemNoStop [0]{.\EOS\space}%
\providecommand \EOS [0]{\spacefactor3000\relax}%
\providecommand \BibitemShut  [1]{\csname bibitem#1\endcsname}%
\let\auto@bib@innerbib\@empty
\bibitem [{\citenamefont {Caves}(1981)}]{Caves1981}%
  \BibitemOpen
  \bibfield  {author} {\bibinfo {author} {\bibfnamefont {C.~M.}\ \bibnamefont
  {Caves}},\ }\bibfield  {title} {\enquote {\bibinfo {title}
  {{Quantum-mechanical noise in an interferometer}},}\ }\href {\doibase
  10.1103/PhysRevD.23.1693} {\bibfield  {journal} {\bibinfo  {journal} {Phys.
  Rev. D}\ }\textbf {\bibinfo {volume} {23}},\ \bibinfo {pages} {1693}
  (\bibinfo {year} {1981})}\BibitemShut {NoStop}%
\bibitem [{\citenamefont {Holland}\ and\ \citenamefont
  {Burnett}(1993)}]{Holland1993}%
  \BibitemOpen
  \bibfield  {author} {\bibinfo {author} {\bibfnamefont {M.~J.}\ \bibnamefont
  {Holland}}\ and\ \bibinfo {author} {\bibfnamefont {K.}~\bibnamefont
  {Burnett}},\ }\bibfield  {title} {\enquote {\bibinfo {title}
  {{Interferometric detection of optical phase shifts at the Heisenberg
  limit}},}\ }\href {\doibase 10.1103/PhysRevLett.71.1355} {\bibfield
  {journal} {\bibinfo  {journal} {Phys. Rev. Lett.}\ }\textbf {\bibinfo
  {volume} {71}},\ \bibinfo {pages} {1355} (\bibinfo {year}
  {1993})}\BibitemShut {NoStop}%
\bibitem [{\citenamefont {Bollinger}\ \emph {et~al.}(1996)\citenamefont
  {Bollinger}, \citenamefont {Itano}, \citenamefont {Wineland},\ and\
  \citenamefont {Heinzen}}]{Bollinger1996}%
  \BibitemOpen
  \bibfield  {author} {\bibinfo {author} {\bibfnamefont {J.~J.}\ \bibnamefont
  {Bollinger}}, \bibinfo {author} {\bibfnamefont {W.~M.}\ \bibnamefont
  {Itano}}, \bibinfo {author} {\bibfnamefont {D.~J.}\ \bibnamefont {Wineland}},
  \ and\ \bibinfo {author} {\bibfnamefont {D.~J.}\ \bibnamefont {Heinzen}},\
  }\bibfield  {title} {\enquote {\bibinfo {title} {Optimal frequency
  measurements with maximally correlated states},}\ }\href {\doibase
  10.1103/PhysRevA.54.R4649} {\bibfield  {journal} {\bibinfo  {journal} {Phys.
  Rev. A}\ }\textbf {\bibinfo {volume} {54}},\ \bibinfo {pages} {R4649}
  (\bibinfo {year} {1996})}\BibitemShut {NoStop}%
\bibitem [{\citenamefont {McKenzie}\ \emph {et~al.}(2002)\citenamefont
  {McKenzie}, \citenamefont {Shaddock}, \citenamefont {McClelland},
  \citenamefont {Buchler},\ and\ \citenamefont {Lam}}]{McKenzie2002}%
  \BibitemOpen
  \bibfield  {author} {\bibinfo {author} {\bibfnamefont {K.}~\bibnamefont
  {McKenzie}}, \bibinfo {author} {\bibfnamefont {D.~A.}\ \bibnamefont
  {Shaddock}}, \bibinfo {author} {\bibfnamefont {D.~E.}\ \bibnamefont
  {McClelland}}, \bibinfo {author} {\bibfnamefont {B.~C.}\ \bibnamefont
  {Buchler}}, \ and\ \bibinfo {author} {\bibfnamefont {P.~K.}\ \bibnamefont
  {Lam}},\ }\bibfield  {title} {\enquote {\bibinfo {title} {Experimental
  demonstration of a squeezing-enhanced power-recycled michelson interferometer
  for gravitational wave detection},}\ }\href {\doibase
  10.1103/PhysRevLett.88.231102} {\bibfield  {journal} {\bibinfo  {journal}
  {Phys. Rev. Lett.}\ }\textbf {\bibinfo {volume} {88}},\ \bibinfo {pages}
  {231102} (\bibinfo {year} {2002})}\BibitemShut {NoStop}%
\bibitem [{\citenamefont {Giovannetti}\ \emph {et~al.}(2011)\citenamefont
  {Giovannetti}, \citenamefont {Lloyd},\ and\ \citenamefont
  {Maccone}}]{GiovannettiNatPhot}%
  \BibitemOpen
  \bibfield  {author} {\bibinfo {author} {\bibfnamefont {V.}~\bibnamefont
  {Giovannetti}}, \bibinfo {author} {\bibfnamefont {S.}~\bibnamefont {Lloyd}},
  \ and\ \bibinfo {author} {\bibfnamefont {L.}~\bibnamefont {Maccone}},\
  }\bibfield  {title} {\enquote {\bibinfo {title} {{Advances in quantum
  metrology}},}\ }\href {\doibase 10.1038/nphoton.2011.35} {\bibfield
  {journal} {\bibinfo  {journal} {Nat. Photonics}\ }\textbf {\bibinfo {volume}
  {5}},\ \bibinfo {pages} {222} (\bibinfo {year} {2011})}\BibitemShut {NoStop}%
\bibitem [{\citenamefont {Huelga}\ \emph {et~al.}(1997)\citenamefont {Huelga},
  \citenamefont {Macchiavello}, \citenamefont {Pellizzari}, \citenamefont
  {Ekert}, \citenamefont {Plenio},\ and\ \citenamefont {Cirac}}]{Huelga97}%
  \BibitemOpen
  \bibfield  {author} {\bibinfo {author} {\bibfnamefont {S.~F.}\ \bibnamefont
  {Huelga}}, \bibinfo {author} {\bibfnamefont {C.}~\bibnamefont
  {Macchiavello}}, \bibinfo {author} {\bibfnamefont {T.}~\bibnamefont
  {Pellizzari}}, \bibinfo {author} {\bibfnamefont {A.~K.}\ \bibnamefont
  {Ekert}}, \bibinfo {author} {\bibfnamefont {M.~B.}\ \bibnamefont {Plenio}}, \
  and\ \bibinfo {author} {\bibfnamefont {J.~I.}\ \bibnamefont {Cirac}},\
  }\bibfield  {title} {\enquote {\bibinfo {title} {{Improvement of Frequency
  Standards with Quantum Entanglement}},}\ }\href {\doibase
  10.1103/PhysRevLett.79.3865} {\bibfield  {journal} {\bibinfo  {journal}
  {Phys. Rev. Lett.}\ }\textbf {\bibinfo {volume} {79}},\ \bibinfo {pages}
  {3865} (\bibinfo {year} {1997})}\BibitemShut {NoStop}%
\bibitem [{\citenamefont {Escher}\ \emph {et~al.}(2011)\citenamefont {Escher},
  \citenamefont {{de Matos Filho}},\ and\ \citenamefont
  {Davidovich}}]{EscherNatPhys}%
  \BibitemOpen
  \bibfield  {author} {\bibinfo {author} {\bibfnamefont {B.~M.}\ \bibnamefont
  {Escher}}, \bibinfo {author} {\bibfnamefont {R.~L.}\ \bibnamefont {{de Matos
  Filho}}}, \ and\ \bibinfo {author} {\bibfnamefont {L.}~\bibnamefont
  {Davidovich}},\ }\bibfield  {title} {\enquote {\bibinfo {title} {{General
  framework for estimating the ultimate precision limit in noisy
  quantum-enhanced metrology}},}\ }\href {\doibase 10.1038/nphys1958}
  {\bibfield  {journal} {\bibinfo  {journal} {Nat. Phys.}\ }\textbf {\bibinfo
  {volume} {7}},\ \bibinfo {pages} {406} (\bibinfo {year} {2011})}\BibitemShut
  {NoStop}%
\bibitem [{\citenamefont {Demkowicz-Dobrza{\'{n}}ski}\ \emph
  {et~al.}(2012)\citenamefont {Demkowicz-Dobrza{\'{n}}ski}, \citenamefont
  {Ko{\l}ody{\'{n}}ski},\ and\ \citenamefont {Gu{\c
  t}\u{a}}}]{KolodynskyNatComm}%
  \BibitemOpen
  \bibfield  {author} {\bibinfo {author} {\bibfnamefont {R.}~\bibnamefont
  {Demkowicz-Dobrza{\'{n}}ski}}, \bibinfo {author} {\bibfnamefont
  {J.}~\bibnamefont {Ko{\l}ody{\'{n}}ski}}, \ and\ \bibinfo {author}
  {\bibfnamefont {M.}~\bibnamefont {Gu{\c t}\u{a}}},\ }\bibfield  {title}
  {\enquote {\bibinfo {title} {{The elusive Heisenberg limit in
  quantum-enhanced metrology}},}\ }\href {\doibase 10.1038/ncomms2067}
  {\bibfield  {journal} {\bibinfo  {journal} {Nat. Commun.}\ }\textbf {\bibinfo
  {volume} {3}},\ \bibinfo {pages} {1063} (\bibinfo {year} {2012})}\BibitemShut
  {NoStop}%
\bibitem [{\citenamefont {Ko{\l}ody{\'{n}}ski}\ and\ \citenamefont
  {Demkowicz-Dobrza{\'{n}}ski}(2013)}]{Koodynski2013}%
  \BibitemOpen
  \bibfield  {author} {\bibinfo {author} {\bibfnamefont {J.}~\bibnamefont
  {Ko{\l}ody{\'{n}}ski}}\ and\ \bibinfo {author} {\bibfnamefont
  {R.}~\bibnamefont {Demkowicz-Dobrza{\'{n}}ski}},\ }\bibfield  {title}
  {\enquote {\bibinfo {title} {{Efficient tools for quantum metrology with
  uncorrelated noise}},}\ }\href {\doibase 10.1088/1367-2630/15/7/073043}
  {\bibfield  {journal} {\bibinfo  {journal} {New J. Phys.}\ }\textbf {\bibinfo
  {volume} {15}},\ \bibinfo {pages} {073043} (\bibinfo {year}
  {2013})}\BibitemShut {NoStop}%
\bibitem [{\citenamefont {Matsuzaki}\ \emph {et~al.}(2011)\citenamefont
  {Matsuzaki}, \citenamefont {Benjamin},\ and\ \citenamefont
  {Fitzsimons}}]{Matsuzaki2011}%
  \BibitemOpen
  \bibfield  {author} {\bibinfo {author} {\bibfnamefont {Y.}~\bibnamefont
  {Matsuzaki}}, \bibinfo {author} {\bibfnamefont {S.~C.}\ \bibnamefont
  {Benjamin}}, \ and\ \bibinfo {author} {\bibfnamefont {J.~F.}\ \bibnamefont
  {Fitzsimons}},\ }\bibfield  {title} {\enquote {\bibinfo {title} {{Magnetic
  field sensing beyond the standard quantum limit under the effect of
  decoherence}},}\ }\href {\doibase 10.1103/PhysRevA.84.012103} {\bibfield
  {journal} {\bibinfo  {journal} {Phys. Rev. A}\ }\textbf {\bibinfo {volume}
  {84}},\ \bibinfo {pages} {012103} (\bibinfo {year} {2011})}\BibitemShut
  {NoStop}%
\bibitem [{\citenamefont {Chin}\ \emph {et~al.}(2012)\citenamefont {Chin},
  \citenamefont {Huelga},\ and\ \citenamefont {Plenio}}]{Chin12}%
  \BibitemOpen
  \bibfield  {author} {\bibinfo {author} {\bibfnamefont {A.~W.}\ \bibnamefont
  {Chin}}, \bibinfo {author} {\bibfnamefont {S.~F.}\ \bibnamefont {Huelga}}, \
  and\ \bibinfo {author} {\bibfnamefont {M.~B.}\ \bibnamefont {Plenio}},\
  }\bibfield  {title} {\enquote {\bibinfo {title} {{Quantum Metrology in
  Non-Markovian Environments}},}\ }\href {\doibase
  10.1103/PhysRevLett.109.233601} {\bibfield  {journal} {\bibinfo  {journal}
  {Phys. Rev. Lett.}\ }\textbf {\bibinfo {volume} {109}},\ \bibinfo {pages}
  {233601} (\bibinfo {year} {2012})}\BibitemShut {NoStop}%
\bibitem [{\citenamefont {Smirne}\ \emph {et~al.}(2016)\citenamefont {Smirne},
  \citenamefont {Ko{\l}ody{\'{n}}ski}, \citenamefont {Huelga},\ and\
  \citenamefont {Demkowicz-Dobrza{\'{n}}ski}}]{Smirne16}%
  \BibitemOpen
  \bibfield  {author} {\bibinfo {author} {\bibfnamefont {A.}~\bibnamefont
  {Smirne}}, \bibinfo {author} {\bibfnamefont {J.}~\bibnamefont
  {Ko{\l}ody{\'{n}}ski}}, \bibinfo {author} {\bibfnamefont {S.~F.}\
  \bibnamefont {Huelga}}, \ and\ \bibinfo {author} {\bibfnamefont
  {R.}~\bibnamefont {Demkowicz-Dobrza{\'{n}}ski}},\ }\bibfield  {title}
  {\enquote {\bibinfo {title} {{Ultimate Precision Limits for Noisy Frequency
  Estimation}},}\ }\href {\doibase 10.1103/PhysRevLett.116.120801} {\bibfield
  {journal} {\bibinfo  {journal} {Phys. Rev. Lett.}\ }\textbf {\bibinfo
  {volume} {116}},\ \bibinfo {pages} {120801} (\bibinfo {year}
  {2016})}\BibitemShut {NoStop}%
\bibitem [{\citenamefont {Haase}\ \emph
  {et~al.}(2018{\natexlab{a}})\citenamefont {Haase}, \citenamefont {Smirne},
  \citenamefont {Ko{\l}ody{\'{n}}ski}, \citenamefont
  {Demkowicz-Dobrza{\'{n}}ski},\ and\ \citenamefont {Huelga}}]{Haase2017}%
  \BibitemOpen
  \bibfield  {author} {\bibinfo {author} {\bibfnamefont {J.~F.}\ \bibnamefont
  {Haase}}, \bibinfo {author} {\bibfnamefont {A.}~\bibnamefont {Smirne}},
  \bibinfo {author} {\bibfnamefont {J.}~\bibnamefont {Ko{\l}ody{\'{n}}ski}},
  \bibinfo {author} {\bibfnamefont {R.}~\bibnamefont
  {Demkowicz-Dobrza{\'{n}}ski}}, \ and\ \bibinfo {author} {\bibfnamefont
  {S.~F.}\ \bibnamefont {Huelga}},\ }\bibfield  {title} {\enquote {\bibinfo
  {title} {{Fundamental limits to frequency estimation: a comprehensive
  microscopic perspective}},}\ }\href {\doibase 10.1088/1367-2630/aab67f}
  {\bibfield  {journal} {\bibinfo  {journal} {New J. Phys.}\ }\textbf {\bibinfo
  {volume} {20}},\ \bibinfo {pages} {053009} (\bibinfo {year}
  {2018}{\natexlab{a}})}\BibitemShut {NoStop}%
\bibitem [{\citenamefont {G{\'{o}}recka}\ \emph {et~al.}(2018)\citenamefont
  {G{\'{o}}recka}, \citenamefont {Pollock}, \citenamefont {Liuzzo-Scorpo},
  \citenamefont {Nichols}, \citenamefont {Adesso},\ and\ \citenamefont
  {Modi}}]{Gorecka2017}%
  \BibitemOpen
  \bibfield  {author} {\bibinfo {author} {\bibfnamefont {A.}~\bibnamefont
  {G{\'{o}}recka}}, \bibinfo {author} {\bibfnamefont {F.~A.}\ \bibnamefont
  {Pollock}}, \bibinfo {author} {\bibfnamefont {P.}~\bibnamefont
  {Liuzzo-Scorpo}}, \bibinfo {author} {\bibfnamefont {R.}~\bibnamefont
  {Nichols}}, \bibinfo {author} {\bibfnamefont {G.}~\bibnamefont {Adesso}}, \
  and\ \bibinfo {author} {\bibfnamefont {K.}~\bibnamefont {Modi}},\ }\bibfield
  {title} {\enquote {\bibinfo {title} {{Noisy frequency estimation with noisy
  probes}},}\ }\href {\doibase 10.1088/1367-2630/aad4e5} {\bibfield  {journal}
  {\bibinfo  {journal} {New J. Phys.}\ }\textbf {\bibinfo {volume} {20}},\
  \bibinfo {pages} {083008} (\bibinfo {year} {2018})}\BibitemShut {NoStop}%
\bibitem [{\citenamefont {Chaves}\ \emph {et~al.}(2013)\citenamefont {Chaves},
  \citenamefont {Brask}, \citenamefont {Markiewicz}, \citenamefont
  {Ko{\l}ody{\'{n}}ski},\ and\ \citenamefont {Ac{\'{i}}n}}]{Chaves2013}%
  \BibitemOpen
  \bibfield  {author} {\bibinfo {author} {\bibfnamefont {R.}~\bibnamefont
  {Chaves}}, \bibinfo {author} {\bibfnamefont {J.~B.}\ \bibnamefont {Brask}},
  \bibinfo {author} {\bibfnamefont {M.}~\bibnamefont {Markiewicz}}, \bibinfo
  {author} {\bibfnamefont {J.}~\bibnamefont {Ko{\l}ody{\'{n}}ski}}, \ and\
  \bibinfo {author} {\bibfnamefont {A.}~\bibnamefont {Ac{\'{i}}n}},\ }\bibfield
   {title} {\enquote {\bibinfo {title} {{Noisy Metrology beyond the Standard
  Quantum Limit}},}\ }\href {\doibase 10.1103/PhysRevLett.111.120401}
  {\bibfield  {journal} {\bibinfo  {journal} {Phys. Rev. Lett.}\ }\textbf
  {\bibinfo {volume} {111}},\ \bibinfo {pages} {120401} (\bibinfo {year}
  {2013})}\BibitemShut {NoStop}%
\bibitem [{\citenamefont {Brask}\ \emph {et~al.}(2015)\citenamefont {Brask},
  \citenamefont {Chaves},\ and\ \citenamefont {Ko{\l}ody{\'{n}}ski}}]{Brask15}%
  \BibitemOpen
  \bibfield  {author} {\bibinfo {author} {\bibfnamefont {J.~B.}\ \bibnamefont
  {Brask}}, \bibinfo {author} {\bibfnamefont {R.}~\bibnamefont {Chaves}}, \
  and\ \bibinfo {author} {\bibfnamefont {J.}~\bibnamefont
  {Ko{\l}ody{\'{n}}ski}},\ }\bibfield  {title} {\enquote {\bibinfo {title}
  {{Improved Quantum Magnetometry beyond the Standard Quantum Limit}},}\ }\href
  {\doibase 10.1103/PhysRevX.5.031010} {\bibfield  {journal} {\bibinfo
  {journal} {Phys. Rev. X}\ }\textbf {\bibinfo {volume} {5}},\ \bibinfo {pages}
  {031010} (\bibinfo {year} {2015})}\BibitemShut {NoStop}%
\bibitem [{\citenamefont {Sekatski}\ \emph {et~al.}(2016)\citenamefont
  {Sekatski}, \citenamefont {Skotiniotis},\ and\ \citenamefont
  {D{\"{u}}r}}]{Sekatski2015a}%
  \BibitemOpen
  \bibfield  {author} {\bibinfo {author} {\bibfnamefont {P.}~\bibnamefont
  {Sekatski}}, \bibinfo {author} {\bibfnamefont {M.}~\bibnamefont
  {Skotiniotis}}, \ and\ \bibinfo {author} {\bibfnamefont {W.}~\bibnamefont
  {D{\"{u}}r}},\ }\bibfield  {title} {\enquote {\bibinfo {title} {{Dynamical
  decoupling leads to improved scaling in noisy quantum metrology}},}\ }\href
  {\doibase 10.1088/1367-2630/18/7/073034} {\bibfield  {journal} {\bibinfo
  {journal} {New J. Phys.}\ }\textbf {\bibinfo {volume} {18}},\ \bibinfo
  {pages} {073034} (\bibinfo {year} {2016})}\BibitemShut {NoStop}%
\bibitem [{\citenamefont {Kessler}\ \emph {et~al.}(2014)\citenamefont
  {Kessler}, \citenamefont {Lovchinsky}, \citenamefont {Sushkov},\ and\
  \citenamefont {Lukin}}]{Kessler14}%
  \BibitemOpen
  \bibfield  {author} {\bibinfo {author} {\bibfnamefont {E.~M.}\ \bibnamefont
  {Kessler}}, \bibinfo {author} {\bibfnamefont {I.}~\bibnamefont {Lovchinsky}},
  \bibinfo {author} {\bibfnamefont {A.~O.}\ \bibnamefont {Sushkov}}, \ and\
  \bibinfo {author} {\bibfnamefont {M.~D.}\ \bibnamefont {Lukin}},\ }\bibfield
  {title} {\enquote {\bibinfo {title} {{Quantum Error Correction for
  Metrology}},}\ }\href {\doibase 10.1103/PhysRevLett.112.150802} {\bibfield
  {journal} {\bibinfo  {journal} {Phys. Rev. Lett.}\ }\textbf {\bibinfo
  {volume} {112}},\ \bibinfo {pages} {150802} (\bibinfo {year}
  {2014})}\BibitemShut {NoStop}%
\bibitem [{\citenamefont {Arrad}\ \emph {et~al.}(2014)\citenamefont {Arrad},
  \citenamefont {Vinkler}, \citenamefont {Aharonov},\ and\ \citenamefont
  {Retzker}}]{Arrad14}%
  \BibitemOpen
  \bibfield  {author} {\bibinfo {author} {\bibfnamefont {G.}~\bibnamefont
  {Arrad}}, \bibinfo {author} {\bibfnamefont {Y.}~\bibnamefont {Vinkler}},
  \bibinfo {author} {\bibfnamefont {D.}~\bibnamefont {Aharonov}}, \ and\
  \bibinfo {author} {\bibfnamefont {A.}~\bibnamefont {Retzker}},\ }\bibfield
  {title} {\enquote {\bibinfo {title} {{Increasing Sensing Resolution with
  Error Correction}},}\ }\href {\doibase 10.1103/PhysRevLett.112.150801}
  {\bibfield  {journal} {\bibinfo  {journal} {Phys. Rev. Lett.}\ }\textbf
  {\bibinfo {volume} {112}},\ \bibinfo {pages} {150801} (\bibinfo {year}
  {2014})}\BibitemShut {NoStop}%
\bibitem [{\citenamefont {D{\"{u}}r}\ \emph {et~al.}(2014)\citenamefont
  {D{\"{u}}r}, \citenamefont {Skotiniotis}, \citenamefont {Fr{\"{o}}wis},\ and\
  \citenamefont {Kraus}}]{Dur14}%
  \BibitemOpen
  \bibfield  {author} {\bibinfo {author} {\bibfnamefont {W.}~\bibnamefont
  {D{\"{u}}r}}, \bibinfo {author} {\bibfnamefont {M.}~\bibnamefont
  {Skotiniotis}}, \bibinfo {author} {\bibfnamefont {F.}~\bibnamefont
  {Fr{\"{o}}wis}}, \ and\ \bibinfo {author} {\bibfnamefont {B.}~\bibnamefont
  {Kraus}},\ }\bibfield  {title} {\enquote {\bibinfo {title} {{Improved Quantum
  Metrology Using Quantum Error Correction}},}\ }\href {\doibase
  10.1103/PhysRevLett.112.080801} {\bibfield  {journal} {\bibinfo  {journal}
  {Phys. Rev. Lett.}\ }\textbf {\bibinfo {volume} {112}},\ \bibinfo {pages}
  {080801} (\bibinfo {year} {2014})}\BibitemShut {NoStop}%
\bibitem [{\citenamefont {Plenio}\ and\ \citenamefont
  {Huelga}(2016)}]{Plenio2016}%
  \BibitemOpen
  \bibfield  {author} {\bibinfo {author} {\bibfnamefont {M.~B.}\ \bibnamefont
  {Plenio}}\ and\ \bibinfo {author} {\bibfnamefont {S.~F.}\ \bibnamefont
  {Huelga}},\ }\bibfield  {title} {\enquote {\bibinfo {title} {{Sensing in the
  presence of an observed environment}},}\ }\href {\doibase
  10.1103/PhysRevA.93.032123} {\bibfield  {journal} {\bibinfo  {journal} {Phys.
  Rev. A}\ }\textbf {\bibinfo {volume} {93}},\ \bibinfo {pages} {032123}
  (\bibinfo {year} {2016})}\BibitemShut {NoStop}%
\bibitem [{\citenamefont {Gefen}\ \emph {et~al.}(2016)\citenamefont {Gefen},
  \citenamefont {Herrera-Mart{\'{i}}},\ and\ \citenamefont
  {Retzker}}]{Gefen2016}%
  \BibitemOpen
  \bibfield  {author} {\bibinfo {author} {\bibfnamefont {T.}~\bibnamefont
  {Gefen}}, \bibinfo {author} {\bibfnamefont {D.~A.}\ \bibnamefont
  {Herrera-Mart{\'{i}}}}, \ and\ \bibinfo {author} {\bibfnamefont
  {A.}~\bibnamefont {Retzker}},\ }\bibfield  {title} {\enquote {\bibinfo
  {title} {{Parameter estimation with efficient photodetectors}},}\ }\href
  {\doibase 10.1103/PhysRevA.93.032133} {\bibfield  {journal} {\bibinfo
  {journal} {Phys. Rev. A}\ }\textbf {\bibinfo {volume} {93}},\ \bibinfo
  {pages} {032133} (\bibinfo {year} {2016})}\BibitemShut {NoStop}%
\bibitem [{\citenamefont {Layden}\ and\ \citenamefont
  {Cappellaro}(2018)}]{Layden2017}%
  \BibitemOpen
  \bibfield  {author} {\bibinfo {author} {\bibfnamefont {D.}~\bibnamefont
  {Layden}}\ and\ \bibinfo {author} {\bibfnamefont {P.}~\bibnamefont
  {Cappellaro}},\ }\bibfield  {title} {\enquote {\bibinfo {title} {{Spatial
  noise filtering through error correction for quantum sensing}},}\ }\href
  {\doibase 10.1038/s41534-018-0082-2} {\bibfield  {journal} {\bibinfo
  {journal} {npj Quantum Inf.}\ }\textbf {\bibinfo {volume} {4}},\ \bibinfo
  {pages} {30} (\bibinfo {year} {2018})}\BibitemShut {NoStop}%
\bibitem [{\citenamefont {Sekatski}\ \emph {et~al.}(2017)\citenamefont
  {Sekatski}, \citenamefont {Skotiniotis}, \citenamefont
  {Ko{\l{}}ody{\'{n}}ski},\ and\ \citenamefont
  {D{\"{u}}r}}]{Sekatski2017metrologyfulland}%
  \BibitemOpen
  \bibfield  {author} {\bibinfo {author} {\bibfnamefont {P.}~\bibnamefont
  {Sekatski}}, \bibinfo {author} {\bibfnamefont {M.}~\bibnamefont
  {Skotiniotis}}, \bibinfo {author} {\bibfnamefont {J.}~\bibnamefont
  {Ko{\l{}}ody{\'{n}}ski}}, \ and\ \bibinfo {author} {\bibfnamefont
  {W.}~\bibnamefont {D{\"{u}}r}},\ }\bibfield  {title} {\enquote {\bibinfo
  {title} {Quantum metrology with full and fast quantum control},}\ }\href
  {\doibase 10.22331/q-2017-09-06-27} {\bibfield  {journal} {\bibinfo
  {journal} {{Quantum}}\ }\textbf {\bibinfo {volume} {1}},\ \bibinfo {pages}
  {27} (\bibinfo {year} {2017})}\BibitemShut {NoStop}%
\bibitem [{\citenamefont {Matsuzaki}\ and\ \citenamefont
  {Benjamin}(2017)}]{Matsuzaki2017}%
  \BibitemOpen
  \bibfield  {author} {\bibinfo {author} {\bibfnamefont {Y.}~\bibnamefont
  {Matsuzaki}}\ and\ \bibinfo {author} {\bibfnamefont {S.}~\bibnamefont
  {Benjamin}},\ }\bibfield  {title} {\enquote {\bibinfo {title} {Magnetic-field
  sensing with quantum error detection under the effect of energy
  relaxation},}\ }\href {\doibase 10.1103/PhysRevA.95.032303} {\bibfield
  {journal} {\bibinfo  {journal} {Phys. Rev. A}\ }\textbf {\bibinfo {volume}
  {95}},\ \bibinfo {pages} {032303} (\bibinfo {year} {2017})}\BibitemShut
  {NoStop}%
\bibitem [{\citenamefont {Zhou}\ \emph {et~al.}(2018)\citenamefont {Zhou},
  \citenamefont {Zhang}, \citenamefont {Preskill},\ and\ \citenamefont
  {Jiang}}]{Zhou2018}%
  \BibitemOpen
  \bibfield  {author} {\bibinfo {author} {\bibfnamefont {S.}~\bibnamefont
  {Zhou}}, \bibinfo {author} {\bibfnamefont {M.}~\bibnamefont {Zhang}},
  \bibinfo {author} {\bibfnamefont {J.}~\bibnamefont {Preskill}}, \ and\
  \bibinfo {author} {\bibfnamefont {L.}~\bibnamefont {Jiang}},\ }\bibfield
  {title} {\enquote {\bibinfo {title} {{Achieving the Heisenberg limit in
  quantum metrology using quantum error correction}},}\ }\href {\doibase
  10.1038/s41467-017-02510-3} {\bibfield  {journal} {\bibinfo  {journal} {Nat.
  Commun.}\ }\textbf {\bibinfo {volume} {9}},\ \bibinfo {pages} {78} (\bibinfo
  {year} {2018})}\BibitemShut {NoStop}%
\bibitem [{\citenamefont {Wiseman}\ and\ \citenamefont
  {Milburn}(2010)}]{WisemanMilburn}%
  \BibitemOpen
  \bibfield  {author} {\bibinfo {author} {\bibfnamefont {H.~M.}\ \bibnamefont
  {Wiseman}}\ and\ \bibinfo {author} {\bibfnamefont {G.~J.}\ \bibnamefont
  {Milburn}},\ }\href@noop {} {\emph {\bibinfo {title} {{Quantum Measurement
  and Control}}}}\ (\bibinfo  {publisher} {Cambridge University Press},\
  \bibinfo {address} {New York},\ \bibinfo {year} {2010})\BibitemShut {NoStop}%
\bibitem [{\citenamefont {Jacobs}\ and\ \citenamefont
  {Steck}(2006)}]{SteckJacobs}%
  \BibitemOpen
  \bibfield  {author} {\bibinfo {author} {\bibfnamefont {K.}~\bibnamefont
  {Jacobs}}\ and\ \bibinfo {author} {\bibfnamefont {D.~A.}\ \bibnamefont
  {Steck}},\ }\bibfield  {title} {\enquote {\bibinfo {title} {{A
  straightforward introduction to continuous quantum measurement}},}\ }\href
  {\doibase 10.1080/00107510601101934} {\bibfield  {journal} {\bibinfo
  {journal} {Contemp. Phys.}\ }\textbf {\bibinfo {volume} {47}},\ \bibinfo
  {pages} {279} (\bibinfo {year} {2006})}\BibitemShut {NoStop}%
\bibitem [{\citenamefont {Mabuchi}(1996)}]{Mabuchi1996}%
  \BibitemOpen
  \bibfield  {author} {\bibinfo {author} {\bibfnamefont {H.}~\bibnamefont
  {Mabuchi}},\ }\bibfield  {title} {\enquote {\bibinfo {title} {Dynamical
  identification of open quantum systems},}\ }\href
  {http://stacks.iop.org/1355-5111/8/i=6/a=002} {\bibfield  {journal} {\bibinfo
   {journal} {Quant. Semiclass. Opt.}\ }\textbf {\bibinfo {volume} {8}},\
  \bibinfo {pages} {1103} (\bibinfo {year} {1996})}\BibitemShut {NoStop}%
\bibitem [{\citenamefont {Verstraete}\ \emph {et~al.}(2001)\citenamefont
  {Verstraete}, \citenamefont {Doherty},\ and\ \citenamefont
  {Mabuchi}}]{Verstraete2001}%
  \BibitemOpen
  \bibfield  {author} {\bibinfo {author} {\bibfnamefont {F.}~\bibnamefont
  {Verstraete}}, \bibinfo {author} {\bibfnamefont {A.~C.}\ \bibnamefont
  {Doherty}}, \ and\ \bibinfo {author} {\bibfnamefont {H.}~\bibnamefont
  {Mabuchi}},\ }\bibfield  {title} {\enquote {\bibinfo {title} {Sensitivity
  optimization in quantum parameter estimation},}\ }\href {\doibase
  10.1103/PhysRevA.64.032111} {\bibfield  {journal} {\bibinfo  {journal} {Phys.
  Rev. A}\ }\textbf {\bibinfo {volume} {64}},\ \bibinfo {pages} {032111}
  (\bibinfo {year} {2001})}\BibitemShut {NoStop}%
\bibitem [{\citenamefont {Gambetta}\ and\ \citenamefont
  {Wiseman}(2001)}]{Gambetta2001}%
  \BibitemOpen
  \bibfield  {author} {\bibinfo {author} {\bibfnamefont {J.}~\bibnamefont
  {Gambetta}}\ and\ \bibinfo {author} {\bibfnamefont {H.~M.}\ \bibnamefont
  {Wiseman}},\ }\bibfield  {title} {\enquote {\bibinfo {title} {State and
  dynamical parameter estimation for open quantum systems},}\ }\href {\doibase
  10.1103/PhysRevA.64.042105} {\bibfield  {journal} {\bibinfo  {journal} {Phys.
  Rev. A}\ }\textbf {\bibinfo {volume} {64}},\ \bibinfo {pages} {042105}
  (\bibinfo {year} {2001})}\BibitemShut {NoStop}%
\bibitem [{\citenamefont {Chase}\ and\ \citenamefont
  {Geremia}(2009)}]{Chase2009}%
  \BibitemOpen
  \bibfield  {author} {\bibinfo {author} {\bibfnamefont {B.~A.}\ \bibnamefont
  {Chase}}\ and\ \bibinfo {author} {\bibfnamefont {J.~M.}\ \bibnamefont
  {Geremia}},\ }\bibfield  {title} {\enquote {\bibinfo {title} {{Single-shot
  parameter estimation via continuous quantum measurement}},}\ }\href {\doibase
  10.1103/PhysRevA.79.022314} {\bibfield  {journal} {\bibinfo  {journal} {Phys.
  Rev. A}\ }\textbf {\bibinfo {volume} {79}},\ \bibinfo {pages} {022314}
  (\bibinfo {year} {2009})}\BibitemShut {NoStop}%
\bibitem [{\citenamefont {Ralph}\ \emph {et~al.}(2011)\citenamefont {Ralph},
  \citenamefont {Jacobs},\ and\ \citenamefont {Hill}}]{Ralph2011a}%
  \BibitemOpen
  \bibfield  {author} {\bibinfo {author} {\bibfnamefont {J.~F.}\ \bibnamefont
  {Ralph}}, \bibinfo {author} {\bibfnamefont {K.}~\bibnamefont {Jacobs}}, \
  and\ \bibinfo {author} {\bibfnamefont {C.~D.}\ \bibnamefont {Hill}},\
  }\bibfield  {title} {\enquote {\bibinfo {title} {{Frequency tracking and
  parameter estimation for robust quantum state estimation}},}\ }\href
  {\doibase 10.1103/PhysRevA.84.052119} {\bibfield  {journal} {\bibinfo
  {journal} {Phys. Rev. A}\ }\textbf {\bibinfo {volume} {84}},\ \bibinfo
  {pages} {052119} (\bibinfo {year} {2011})}\BibitemShut {NoStop}%
\bibitem [{\citenamefont {Six}\ \emph {et~al.}(2015)\citenamefont {Six},
  \citenamefont {Campagne-Ibarcq}, \citenamefont {Bretheau}, \citenamefont
  {Huard},\ and\ \citenamefont {Rouchon}}]{Six2015}%
  \BibitemOpen
  \bibfield  {author} {\bibinfo {author} {\bibfnamefont {P.}~\bibnamefont
  {Six}}, \bibinfo {author} {\bibfnamefont {P.}~\bibnamefont
  {Campagne-Ibarcq}}, \bibinfo {author} {\bibfnamefont {L.}~\bibnamefont
  {Bretheau}}, \bibinfo {author} {\bibfnamefont {B.}~\bibnamefont {Huard}}, \
  and\ \bibinfo {author} {\bibfnamefont {P.}~\bibnamefont {Rouchon}},\
  }\bibfield  {title} {\enquote {\bibinfo {title} {{Parameter estimation from
  measurements along quantum trajectories}},}\ }in\ \href {\doibase
  10.1109/CDC.2015.7403443} {\emph {\bibinfo {booktitle} {2015 54th IEEE Conf.
  Decis. Control}}},\ \bibinfo {series and number} {\bibinfo {number} {Cdc}}\
  (\bibinfo  {publisher} {IEEE},\ \bibinfo {year} {2015})\ p.\ \bibinfo {pages}
  {7742}\BibitemShut {NoStop}%
\bibitem [{\citenamefont {Cortez}\ \emph {et~al.}(2017)\citenamefont {Cortez},
  \citenamefont {Chantasri}, \citenamefont {Garc{\'{i}}a-Pintos}, \citenamefont
  {Dressel},\ and\ \citenamefont {Jordan}}]{Cortez2017}%
  \BibitemOpen
  \bibfield  {author} {\bibinfo {author} {\bibfnamefont {L.}~\bibnamefont
  {Cortez}}, \bibinfo {author} {\bibfnamefont {A.}~\bibnamefont {Chantasri}},
  \bibinfo {author} {\bibfnamefont {L.~P.}\ \bibnamefont
  {Garc{\'{i}}a-Pintos}}, \bibinfo {author} {\bibfnamefont {J.}~\bibnamefont
  {Dressel}}, \ and\ \bibinfo {author} {\bibfnamefont {A.~N.}\ \bibnamefont
  {Jordan}},\ }\bibfield  {title} {\enquote {\bibinfo {title} {{Rapid
  estimation of drifting parameters in continuously measured quantum
  systems}},}\ }\href {\doibase 10.1103/PhysRevA.95.012314} {\bibfield
  {journal} {\bibinfo  {journal} {Phys. Rev. A}\ }\textbf {\bibinfo {volume}
  {95}},\ \bibinfo {pages} {012314} (\bibinfo {year} {2017})}\BibitemShut
  {NoStop}%
\bibitem [{\citenamefont {Ralph}\ \emph {et~al.}(2017)\citenamefont {Ralph},
  \citenamefont {Maskell},\ and\ \citenamefont {Jacobs}}]{Ralph2017}%
  \BibitemOpen
  \bibfield  {author} {\bibinfo {author} {\bibfnamefont {J.~F.}\ \bibnamefont
  {Ralph}}, \bibinfo {author} {\bibfnamefont {S.}~\bibnamefont {Maskell}}, \
  and\ \bibinfo {author} {\bibfnamefont {K.}~\bibnamefont {Jacobs}},\
  }\bibfield  {title} {\enquote {\bibinfo {title} {{Multiparameter estimation
  along quantum trajectories with sequential Monte Carlo methods}},}\ }\href
  {\doibase 10.1103/PhysRevA.96.052306} {\bibfield  {journal} {\bibinfo
  {journal} {Phys. Rev. A}\ }\textbf {\bibinfo {volume} {96}},\ \bibinfo
  {pages} {052306} (\bibinfo {year} {2017})}\BibitemShut {NoStop}%
\bibitem [{\citenamefont {Tsang}(2009)}]{Tsang2009b}%
  \BibitemOpen
  \bibfield  {author} {\bibinfo {author} {\bibfnamefont {M.}~\bibnamefont
  {Tsang}},\ }\bibfield  {title} {\enquote {\bibinfo {title} {{Optimal waveform
  estimation for classical and quantum systems via time-symmetric
  smoothing}},}\ }\href {\doibase 10.1103/PhysRevA.80.033840} {\bibfield
  {journal} {\bibinfo  {journal} {Phys. Rev. A}\ }\textbf {\bibinfo {volume}
  {80}},\ \bibinfo {pages} {033840} (\bibinfo {year} {2009})}\BibitemShut
  {NoStop}%
\bibitem [{\citenamefont {Tsang}\ \emph {et~al.}(2011)\citenamefont {Tsang},
  \citenamefont {Wiseman},\ and\ \citenamefont {Caves}}]{Tsang2011}%
  \BibitemOpen
  \bibfield  {author} {\bibinfo {author} {\bibfnamefont {M.}~\bibnamefont
  {Tsang}}, \bibinfo {author} {\bibfnamefont {H.~M.}\ \bibnamefont {Wiseman}},
  \ and\ \bibinfo {author} {\bibfnamefont {C.~M.}\ \bibnamefont {Caves}},\
  }\bibfield  {title} {\enquote {\bibinfo {title} {Fundamental quantum limit to
  waveform estimation},}\ }\href {\doibase 10.1103/PhysRevLett.106.090401}
  {\bibfield  {journal} {\bibinfo  {journal} {Phys. Rev. Lett.}\ }\textbf
  {\bibinfo {volume} {106}},\ \bibinfo {pages} {090401} (\bibinfo {year}
  {2011})}\BibitemShut {NoStop}%
\bibitem [{\citenamefont {Ng}\ \emph {et~al.}(2016)\citenamefont {Ng},
  \citenamefont {Ang}, \citenamefont {Wheatley}, \citenamefont {Yonezawa},
  \citenamefont {Furusawa}, \citenamefont {Huntington},\ and\ \citenamefont
  {Tsang}}]{Ng2016}%
  \BibitemOpen
  \bibfield  {author} {\bibinfo {author} {\bibfnamefont {S.}~\bibnamefont
  {Ng}}, \bibinfo {author} {\bibfnamefont {S.~Z.}\ \bibnamefont {Ang}},
  \bibinfo {author} {\bibfnamefont {T.~A.}\ \bibnamefont {Wheatley}}, \bibinfo
  {author} {\bibfnamefont {H.}~\bibnamefont {Yonezawa}}, \bibinfo {author}
  {\bibfnamefont {A.}~\bibnamefont {Furusawa}}, \bibinfo {author}
  {\bibfnamefont {E.~H.}\ \bibnamefont {Huntington}}, \ and\ \bibinfo {author}
  {\bibfnamefont {M.}~\bibnamefont {Tsang}},\ }\bibfield  {title} {\enquote
  {\bibinfo {title} {{Spectrum analysis with quantum dynamical systems}},}\
  }\href {\doibase 10.1103/PhysRevA.93.042121} {\bibfield  {journal} {\bibinfo
  {journal} {Phys. Rev. A}\ }\textbf {\bibinfo {volume} {93}},\ \bibinfo
  {pages} {042121} (\bibinfo {year} {2016})}\BibitemShut {NoStop}%
\bibitem [{\citenamefont {Guţă}(2011)}]{Guta2011}%
  \BibitemOpen
  \bibfield  {author} {\bibinfo {author} {\bibfnamefont {M.}~\bibnamefont
  {Guţă}},\ }\bibfield  {title} {\enquote {\bibinfo {title} {{Fisher
  information and asymptotic normality in system identification for quantum
  Markov chains}},}\ }\href {\doibase 10.1103/PhysRevA.83.062324} {\bibfield
  {journal} {\bibinfo  {journal} {Phys. Rev. A}\ }\textbf {\bibinfo {volume}
  {83}},\ \bibinfo {pages} {062324} (\bibinfo {year} {2011})}\BibitemShut
  {NoStop}%
\bibitem [{\citenamefont {Gammelmark}\ and\ \citenamefont
  {M{\o}lmer}(2013)}]{GammelmarkCRB}%
  \BibitemOpen
  \bibfield  {author} {\bibinfo {author} {\bibfnamefont {S.}~\bibnamefont
  {Gammelmark}}\ and\ \bibinfo {author} {\bibfnamefont {K.}~\bibnamefont
  {M{\o}lmer}},\ }\bibfield  {title} {\enquote {\bibinfo {title} {{Bayesian
  parameter inference from continuously monitored quantum systems}},}\ }\href
  {\doibase 10.1103/PhysRevA.87.032115} {\bibfield  {journal} {\bibinfo
  {journal} {Phys. Rev. A}\ }\textbf {\bibinfo {volume} {87}},\ \bibinfo
  {pages} {032115} (\bibinfo {year} {2013})}\BibitemShut {NoStop}%
\bibitem [{\citenamefont {Gammelmark}\ and\ \citenamefont
  {M{\o}lmer}(2014)}]{GammelmarkQCRB}%
  \BibitemOpen
  \bibfield  {author} {\bibinfo {author} {\bibfnamefont {S.}~\bibnamefont
  {Gammelmark}}\ and\ \bibinfo {author} {\bibfnamefont {K.}~\bibnamefont
  {M{\o}lmer}},\ }\bibfield  {title} {\enquote {\bibinfo {title} {{Fisher
  Information and the Quantum Cram{\'{e}}r-Rao Sensitivity Limit of Continuous
  Measurements}},}\ }\href {\doibase 10.1103/PhysRevLett.112.170401} {\bibfield
   {journal} {\bibinfo  {journal} {Phys. Rev. Lett.}\ }\textbf {\bibinfo
  {volume} {112}},\ \bibinfo {pages} {170401} (\bibinfo {year}
  {2014})}\BibitemShut {NoStop}%
\bibitem [{\citenamefont {Macieszczak}\ \emph {et~al.}(2016)\citenamefont
  {Macieszczak}, \citenamefont {Guţă}, \citenamefont {Lesanovsky},\ and\
  \citenamefont {Garrahan}}]{Macieszczak2016}%
  \BibitemOpen
  \bibfield  {author} {\bibinfo {author} {\bibfnamefont {K.}~\bibnamefont
  {Macieszczak}}, \bibinfo {author} {\bibfnamefont {M.}~\bibnamefont {Guţă}},
  \bibinfo {author} {\bibfnamefont {I.}~\bibnamefont {Lesanovsky}}, \ and\
  \bibinfo {author} {\bibfnamefont {J.~P.}\ \bibnamefont {Garrahan}},\
  }\bibfield  {title} {\enquote {\bibinfo {title} {{Dynamical phase transitions
  as a resource for quantum enhanced metrology}},}\ }\href {\doibase
  10.1103/PhysRevA.93.022103} {\bibfield  {journal} {\bibinfo  {journal} {Phys.
  Rev. A}\ }\textbf {\bibinfo {volume} {93}},\ \bibinfo {pages} {022103}
  (\bibinfo {year} {2016})}\BibitemShut {NoStop}%
\bibitem [{\citenamefont {Genoni}(2017)}]{Genoni2017}%
  \BibitemOpen
  \bibfield  {author} {\bibinfo {author} {\bibfnamefont {M.~G.}\ \bibnamefont
  {Genoni}},\ }\bibfield  {title} {\enquote {\bibinfo {title}
  {{Cram{\'{e}}r-Rao bound for time-continuous measurements in linear Gaussian
  quantum systems}},}\ }\href {\doibase 10.1103/PhysRevA.95.012116} {\bibfield
  {journal} {\bibinfo  {journal} {Phys. Rev. A}\ }\textbf {\bibinfo {volume}
  {95}},\ \bibinfo {pages} {012116} (\bibinfo {year} {2017})}\BibitemShut
  {NoStop}%
\bibitem [{\citenamefont {Kiilerich}\ and\ \citenamefont
  {M{\o}lmer}(2014)}]{KiilerichPC}%
  \BibitemOpen
  \bibfield  {author} {\bibinfo {author} {\bibfnamefont {A.~H.}\ \bibnamefont
  {Kiilerich}}\ and\ \bibinfo {author} {\bibfnamefont {K.}~\bibnamefont
  {M{\o}lmer}},\ }\bibfield  {title} {\enquote {\bibinfo {title} {{Estimation
  of atomic interaction parameters by photon counting}},}\ }\href {\doibase
  10.1103/PhysRevA.89.052110} {\bibfield  {journal} {\bibinfo  {journal} {Phys.
  Rev. A}\ }\textbf {\bibinfo {volume} {89}},\ \bibinfo {pages} {052110}
  (\bibinfo {year} {2014})}\BibitemShut {NoStop}%
\bibitem [{\citenamefont {Kiilerich}\ and\ \citenamefont
  {M\o{}lmer}(2015)}]{Kiilerich2015a}%
  \BibitemOpen
  \bibfield  {author} {\bibinfo {author} {\bibfnamefont {A.~H.}\ \bibnamefont
  {Kiilerich}}\ and\ \bibinfo {author} {\bibfnamefont {K.}~\bibnamefont
  {M\o{}lmer}},\ }\bibfield  {title} {\enquote {\bibinfo {title} {Parameter
  estimation by multichannel photon counting},}\ }\href {\doibase
  10.1103/PhysRevA.91.012119} {\bibfield  {journal} {\bibinfo  {journal} {Phys.
  Rev. A}\ }\textbf {\bibinfo {volume} {91}},\ \bibinfo {pages} {012119}
  (\bibinfo {year} {2015})}\BibitemShut {NoStop}%
\bibitem [{\citenamefont {Kiilerich}\ and\ \citenamefont
  {M{\o}lmer}(2016)}]{KiilerichHomodyne}%
  \BibitemOpen
  \bibfield  {author} {\bibinfo {author} {\bibfnamefont {A.~H.}\ \bibnamefont
  {Kiilerich}}\ and\ \bibinfo {author} {\bibfnamefont {K.}~\bibnamefont
  {M{\o}lmer}},\ }\bibfield  {title} {\enquote {\bibinfo {title} {{Bayesian
  parameter estimation by continuous homodyne detection}},}\ }\href {\doibase
  10.1103/PhysRevA.94.032103} {\bibfield  {journal} {\bibinfo  {journal} {Phys.
  Rev. A}\ }\textbf {\bibinfo {volume} {94}},\ \bibinfo {pages} {032103}
  (\bibinfo {year} {2016})}\BibitemShut {NoStop}%
\bibitem [{\citenamefont {Albarelli}\ \emph {et~al.}(2017)\citenamefont
  {Albarelli}, \citenamefont {Rossi}, \citenamefont {Paris},\ and\
  \citenamefont {Genoni}}]{Albarelli2017a}%
  \BibitemOpen
  \bibfield  {author} {\bibinfo {author} {\bibfnamefont {F.}~\bibnamefont
  {Albarelli}}, \bibinfo {author} {\bibfnamefont {M.~A.~C.}\ \bibnamefont
  {Rossi}}, \bibinfo {author} {\bibfnamefont {M.~G.~A.}\ \bibnamefont {Paris}},
  \ and\ \bibinfo {author} {\bibfnamefont {M.~G.}\ \bibnamefont {Genoni}},\
  }\bibfield  {title} {\enquote {\bibinfo {title} {{Ultimate limits for quantum
  magnetometry via time-continuous measurements}},}\ }\href {\doibase
  10.1088/1367-2630/aa9840} {\bibfield  {journal} {\bibinfo  {journal} {New J.
  Phys.}\ }\textbf {\bibinfo {volume} {19}},\ \bibinfo {pages} {123011}
  (\bibinfo {year} {2017})}\BibitemShut {NoStop}%
\bibitem [{\citenamefont {Geremia}\ \emph {et~al.}(2003)\citenamefont
  {Geremia}, \citenamefont {Stockton}, \citenamefont {Doherty},\ and\
  \citenamefont {Mabuchi}}]{Geremia2003}%
  \BibitemOpen
  \bibfield  {author} {\bibinfo {author} {\bibfnamefont {J.~M.}\ \bibnamefont
  {Geremia}}, \bibinfo {author} {\bibfnamefont {J.~K.}\ \bibnamefont
  {Stockton}}, \bibinfo {author} {\bibfnamefont {A.~C.}\ \bibnamefont
  {Doherty}}, \ and\ \bibinfo {author} {\bibfnamefont {H.}~\bibnamefont
  {Mabuchi}},\ }\bibfield  {title} {\enquote {\bibinfo {title} {{Quantum Kalman
  Filtering and the Heisenberg Limit in Atomic Magnetometry}},}\ }\href
  {\doibase 10.1103/PhysRevLett.91.250801} {\bibfield  {journal} {\bibinfo
  {journal} {Phys. Rev. Lett.}\ }\textbf {\bibinfo {volume} {91}},\ \bibinfo
  {pages} {250801} (\bibinfo {year} {2003})}\BibitemShut {NoStop}%
\bibitem [{\citenamefont {M{\o}lmer}\ and\ \citenamefont
  {Madsen}(2004)}]{Molmer2004}%
  \BibitemOpen
  \bibfield  {author} {\bibinfo {author} {\bibfnamefont {K.}~\bibnamefont
  {M{\o}lmer}}\ and\ \bibinfo {author} {\bibfnamefont {L.~B.}\ \bibnamefont
  {Madsen}},\ }\bibfield  {title} {\enquote {\bibinfo {title} {{Estimation of a
  classical parameter with Gaussian probes: Magnetometry with collective atomic
  spins}},}\ }\href {\doibase 10.1103/PhysRevA.70.052102} {\bibfield  {journal}
  {\bibinfo  {journal} {Phys. Rev. A}\ }\textbf {\bibinfo {volume} {70}},\
  \bibinfo {pages} {052102} (\bibinfo {year} {2004})}\BibitemShut {NoStop}%
\bibitem [{\citenamefont {Catana}\ and\ \citenamefont
  {Guţă}(2014)}]{Catana2014}%
  \BibitemOpen
  \bibfield  {author} {\bibinfo {author} {\bibfnamefont {C.}~\bibnamefont
  {Catana}}\ and\ \bibinfo {author} {\bibfnamefont {M.}~\bibnamefont
  {Guţă}},\ }\bibfield  {title} {\enquote {\bibinfo {title} {Heisenberg
  versus standard scaling in quantum metrology with markov generated states and
  monitored environment},}\ }\href {\doibase 10.1103/PhysRevA.90.012330}
  {\bibfield  {journal} {\bibinfo  {journal} {Phys. Rev. A}\ }\textbf {\bibinfo
  {volume} {90}},\ \bibinfo {pages} {012330} (\bibinfo {year}
  {2014})}\BibitemShut {NoStop}%
\bibitem [{\citenamefont {Braunstein}\ and\ \citenamefont
  {Caves}(1994)}]{CavesBraunstein}%
  \BibitemOpen
  \bibfield  {author} {\bibinfo {author} {\bibfnamefont {S.~L.}\ \bibnamefont
  {Braunstein}}\ and\ \bibinfo {author} {\bibfnamefont {C.~M.}\ \bibnamefont
  {Caves}},\ }\bibfield  {title} {\enquote {\bibinfo {title} {{Statistical
  distance and the geometry of quantum states}},}\ }\href {\doibase
  10.1103/PhysRevLett.72.3439} {\bibfield  {journal} {\bibinfo  {journal}
  {Phys. Rev. Lett.}\ }\textbf {\bibinfo {volume} {72}},\ \bibinfo {pages}
  {3439} (\bibinfo {year} {1994})}\BibitemShut {NoStop}%
\bibitem [{\citenamefont {Rouchon}\ and\ \citenamefont
  {Ralph}(2015)}]{Rouchon2015}%
  \BibitemOpen
  \bibfield  {author} {\bibinfo {author} {\bibfnamefont {P.}~\bibnamefont
  {Rouchon}}\ and\ \bibinfo {author} {\bibfnamefont {J.~F.}\ \bibnamefont
  {Ralph}},\ }\bibfield  {title} {\enquote {\bibinfo {title} {{Efficient
  quantum filtering for quantum feedback control}},}\ }\href {\doibase
  10.1103/PhysRevA.91.012118} {\bibfield  {journal} {\bibinfo  {journal} {Phys.
  Rev. A}\ }\textbf {\bibinfo {volume} {91}},\ \bibinfo {pages} {012118}
  (\bibinfo {year} {2015})}\BibitemShut {NoStop}%
\bibitem [{\citenamefont {Combes}\ \emph {et~al.}(2014)\citenamefont {Combes},
  \citenamefont {Ferrie}, \citenamefont {Jiang},\ and\ \citenamefont
  {Caves}}]{Combes2014}%
  \BibitemOpen
  \bibfield  {author} {\bibinfo {author} {\bibfnamefont {J.}~\bibnamefont
  {Combes}}, \bibinfo {author} {\bibfnamefont {C.}~\bibnamefont {Ferrie}},
  \bibinfo {author} {\bibfnamefont {Z.}~\bibnamefont {Jiang}}, \ and\ \bibinfo
  {author} {\bibfnamefont {C.~M.}\ \bibnamefont {Caves}},\ }\bibfield  {title}
  {\enquote {\bibinfo {title} {{Quantum limits on postselected, probabilistic
  quantum metrology}},}\ }\href {\doibase 10.1103/PhysRevA.89.052117}
  {\bibfield  {journal} {\bibinfo  {journal} {Phys. Rev. A}\ }\textbf {\bibinfo
  {volume} {89}},\ \bibinfo {pages} {052117} (\bibinfo {year}
  {2014})}\BibitemShut {NoStop}%
\bibitem [{\citenamefont {Zhang}\ \emph {et~al.}(2015)\citenamefont {Zhang},
  \citenamefont {Datta},\ and\ \citenamefont {Walmsley}}]{Zhang2015}%
  \BibitemOpen
  \bibfield  {author} {\bibinfo {author} {\bibfnamefont {L.}~\bibnamefont
  {Zhang}}, \bibinfo {author} {\bibfnamefont {A.}~\bibnamefont {Datta}}, \ and\
  \bibinfo {author} {\bibfnamefont {I.~A.}\ \bibnamefont {Walmsley}},\
  }\bibfield  {title} {\enquote {\bibinfo {title} {{Precision Metrology Using
  Weak Measurements}},}\ }\href {\doibase 10.1103/PhysRevLett.114.210801}
  {\bibfield  {journal} {\bibinfo  {journal} {Phys. Rev. Lett.}\ }\textbf
  {\bibinfo {volume} {114}},\ \bibinfo {pages} {210801} (\bibinfo {year}
  {2015})}\BibitemShut {NoStop}%
\bibitem [{\citenamefont {Alves}\ \emph {et~al.}(2015)\citenamefont {Alves},
  \citenamefont {Escher}, \citenamefont {{de Matos Filho}}, \citenamefont
  {Zagury},\ and\ \citenamefont {Davidovich}}]{Alves2015}%
  \BibitemOpen
  \bibfield  {author} {\bibinfo {author} {\bibfnamefont {G.~B.}\ \bibnamefont
  {Alves}}, \bibinfo {author} {\bibfnamefont {B.~M.}\ \bibnamefont {Escher}},
  \bibinfo {author} {\bibfnamefont {R.~L.}\ \bibnamefont {{de Matos Filho}}},
  \bibinfo {author} {\bibfnamefont {N.}~\bibnamefont {Zagury}}, \ and\ \bibinfo
  {author} {\bibfnamefont {L.}~\bibnamefont {Davidovich}},\ }\bibfield  {title}
  {\enquote {\bibinfo {title} {{Weak-value amplification as an optimal
  metrological protocol}},}\ }\href {\doibase 10.1103/PhysRevA.91.062107}
  {\bibfield  {journal} {\bibinfo  {journal} {Phys. Rev. A}\ }\textbf {\bibinfo
  {volume} {91}},\ \bibinfo {pages} {062107} (\bibinfo {year}
  {2015})}\BibitemShut {NoStop}%
\bibitem [{\citenamefont {Alipour}\ and\ \citenamefont
  {Rezakhani}(2015)}]{Alipour2015}%
  \BibitemOpen
  \bibfield  {author} {\bibinfo {author} {\bibfnamefont {S.}~\bibnamefont
  {Alipour}}\ and\ \bibinfo {author} {\bibfnamefont {A.~T.}\ \bibnamefont
  {Rezakhani}},\ }\bibfield  {title} {\enquote {\bibinfo {title} {Extended
  convexity of quantum fisher information in quantum metrology},}\ }\href
  {\doibase 10.1103/PhysRevA.91.042104} {\bibfield  {journal} {\bibinfo
  {journal} {Phys. Rev. A}\ }\textbf {\bibinfo {volume} {91}},\ \bibinfo
  {pages} {042104} (\bibinfo {year} {2015})}\BibitemShut {NoStop}%
\bibitem [{\citenamefont {Alipour}\ \emph {et~al.}(2014)\citenamefont
  {Alipour}, \citenamefont {Mehboudi},\ and\ \citenamefont
  {Rezakhani}}]{Alipour2014}%
  \BibitemOpen
  \bibfield  {author} {\bibinfo {author} {\bibfnamefont {S.}~\bibnamefont
  {Alipour}}, \bibinfo {author} {\bibfnamefont {M.}~\bibnamefont {Mehboudi}}, \
  and\ \bibinfo {author} {\bibfnamefont {A.~T.}\ \bibnamefont {Rezakhani}},\
  }\bibfield  {title} {\enquote {\bibinfo {title} {{Quantum Metrology in Open
  Systems: Dissipative Cram{\'{e}}r-Rao Bound}},}\ }\href {\doibase
  10.1103/PhysRevLett.112.120405} {\bibfield  {journal} {\bibinfo  {journal}
  {Phys. Rev. Lett.}\ }\textbf {\bibinfo {volume} {112}},\ \bibinfo {pages}
  {120405} (\bibinfo {year} {2014})}\BibitemShut {NoStop}%
\bibitem [{\citenamefont {Beau}\ and\ \citenamefont {del
  Campo}(2017)}]{Beau2017}%
  \BibitemOpen
  \bibfield  {author} {\bibinfo {author} {\bibfnamefont {M.}~\bibnamefont
  {Beau}}\ and\ \bibinfo {author} {\bibfnamefont {A.}~\bibnamefont {del
  Campo}},\ }\bibfield  {title} {\enquote {\bibinfo {title} {{Nonlinear Quantum
  Metrology of Many-Body Open Systems}},}\ }\href {\doibase
  10.1103/PhysRevLett.119.010403} {\bibfield  {journal} {\bibinfo  {journal}
  {Phys. Rev. Lett.}\ }\textbf {\bibinfo {volume} {119}},\ \bibinfo {pages}
  {010403} (\bibinfo {year} {2017})}\BibitemShut {NoStop}%
\bibitem [{\citenamefont {Albarelli}(2018)}]{Albarelli2018thesis}%
  \BibitemOpen
  \bibfield  {author} {\bibinfo {author} {\bibfnamefont {F.}~\bibnamefont
  {Albarelli}},\ }\emph {\bibinfo {title} {{Continuous measurements and
  nonclassicality as resources for quantum technologies}}},\ \href@noop {}
  {\bibinfo {type} {{PhD thesis}}},\ \bibinfo  {school} {Universit{\`{a}} degli
  Studi di Milano} (\bibinfo {year} {2018})\BibitemShut {NoStop}%
\bibitem [{\citenamefont {{\v{S}}afr{\'{a}}nek}(2017)}]{Safranek2017}%
  \BibitemOpen
  \bibfield  {author} {\bibinfo {author} {\bibfnamefont {D.}~\bibnamefont
  {{\v{S}}afr{\'{a}}nek}},\ }\bibfield  {title} {\enquote {\bibinfo {title}
  {{Discontinuities of the quantum Fisher information and the Bures metric}},}\
  }\href {\doibase 10.1103/PhysRevA.95.052320} {\bibfield  {journal} {\bibinfo
  {journal} {Phys. Rev. A}\ }\textbf {\bibinfo {volume} {95}},\ \bibinfo
  {pages} {052320} (\bibinfo {year} {2017})}\BibitemShut {NoStop}%
\bibitem [{\citenamefont {Seveso}\ \emph {et~al.}(2018)\citenamefont {Seveso},
  \citenamefont {Albarelli}, \citenamefont {Genoni},\ and\ \citenamefont
  {Paris}}]{SevesoTBA}%
  \BibitemOpen
  \bibfield  {author} {\bibinfo {author} {\bibfnamefont {L.}~\bibnamefont
  {Seveso}}, \bibinfo {author} {\bibfnamefont {F.}~\bibnamefont {Albarelli}},
  \bibinfo {author} {\bibfnamefont {M.~G.}\ \bibnamefont {Genoni}}, \ and\
  \bibinfo {author} {\bibfnamefont {M.~G.~A.}\ \bibnamefont {Paris}},\
  }\href@noop {} {\bibfield  {journal} {\bibinfo  {journal} {\emph{in
  preparation}}\ } (\bibinfo {year} {2018})}\BibitemShut {NoStop}%
\bibitem [{\citenamefont {Andersson}\ \emph {et~al.}(2007)\citenamefont
  {Andersson}, \citenamefont {Cresser},\ and\ \citenamefont
  {Hall}}]{Andersson2007}%
  \BibitemOpen
  \bibfield  {author} {\bibinfo {author} {\bibfnamefont {E.}~\bibnamefont
  {Andersson}}, \bibinfo {author} {\bibfnamefont {J.~D.}\ \bibnamefont
  {Cresser}}, \ and\ \bibinfo {author} {\bibfnamefont {M.~J.~W.}\ \bibnamefont
  {Hall}},\ }\bibfield  {title} {\enquote {\bibinfo {title} {{Finding the Kraus
  decomposition from a master equation and vice versa}},}\ }\href {\doibase
  10.1080/09500340701352581} {\bibfield  {journal} {\bibinfo  {journal} {J.
  Mod. Opt.}\ }\textbf {\bibinfo {volume} {54}},\ \bibinfo {pages} {1695}
  (\bibinfo {year} {2007})}\BibitemShut {NoStop}%
\bibitem [{\citenamefont {Albarelli}\ \emph {et~al.}(2018)\citenamefont
  {Albarelli}, \citenamefont {Rossi}, \citenamefont {Tamascelli},\ and\
  \citenamefont {Genoni}}]{ContinuousMeasurementFI}%
  \BibitemOpen
  \bibfield  {author} {\bibinfo {author} {\bibfnamefont {F.}~\bibnamefont
  {Albarelli}}, \bibinfo {author} {\bibfnamefont {M.~A.~C.}\ \bibnamefont
  {Rossi}}, \bibinfo {author} {\bibfnamefont {D.}~\bibnamefont {Tamascelli}}, \
  and\ \bibinfo {author} {\bibfnamefont {M.~G.}\ \bibnamefont {Genoni}},\
  }\href {\doibase 10.5281/zenodo.1456660} {\enquote {\bibinfo {title}
  {{C}ontinuous{M}easurement{FI}},}\ } (\bibinfo {year} {2018}),\ \bibinfo
  {note}
  {\url{https://github.com/matteoacrossi/ContinuousMeasurementFI}}\BibitemShut
  {NoStop}%
\bibitem [{\citenamefont {Bezanson}\ \emph {et~al.}(2017)\citenamefont
  {Bezanson}, \citenamefont {Edelman}, \citenamefont {Karpinski},\ and\
  \citenamefont {Shah}}]{julia}%
  \BibitemOpen
  \bibfield  {author} {\bibinfo {author} {\bibfnamefont {J.}~\bibnamefont
  {Bezanson}}, \bibinfo {author} {\bibfnamefont {A.}~\bibnamefont {Edelman}},
  \bibinfo {author} {\bibfnamefont {S.}~\bibnamefont {Karpinski}}, \ and\
  \bibinfo {author} {\bibfnamefont {V.~B.}\ \bibnamefont {Shah}},\ }\bibfield
  {title} {\enquote {\bibinfo {title} {Julia: A fresh approach to numerical
  computing},}\ }\href {\doibase 10.1137/141000671} {\bibfield  {journal}
  {\bibinfo  {journal} {SIAM Rev.}\ }\textbf {\bibinfo {volume} {59}},\
  \bibinfo {pages} {65} (\bibinfo {year} {2017})}\BibitemShut {NoStop}%
\bibitem [{\citenamefont {{Rouchon}}(2014)}]{Rouchon2014}%
  \BibitemOpen
  \bibfield  {author} {\bibinfo {author} {\bibfnamefont {P.}~\bibnamefont
  {{Rouchon}}},\ }\bibfield  {title} {\enquote {\bibinfo {title} {{Models and
  Feedback Stabilization of Open Quantum Systems}},}\ }\href
  {https://arxiv.org/abs/1407.7810} {\bibfield  {journal} {\bibinfo  {journal}
  {arXiv:1407.7810}\ } (\bibinfo {year} {2014})}\BibitemShut {NoStop}%
\bibitem [{\citenamefont {Paris}(2009)}]{MatteoIJQI}%
  \BibitemOpen
  \bibfield  {author} {\bibinfo {author} {\bibfnamefont {M.~G.~A.}\
  \bibnamefont {Paris}},\ }\bibfield  {title} {\enquote {\bibinfo {title}
  {{Quantum Estimation for Quantum Technology}},}\ }\href {\doibase
  10.1142/S0219749909004839} {\bibfield  {journal} {\bibinfo  {journal} {Int.
  J. Quant. Inf.}\ }\textbf {\bibinfo {volume} {07}},\ \bibinfo {pages} {125}
  (\bibinfo {year} {2009})}\BibitemShut {NoStop}%
\bibitem [{\citenamefont {Haase}\ \emph
  {et~al.}(2018{\natexlab{b}})\citenamefont {Haase}, \citenamefont {Smirne},
  \citenamefont {Huelga}, \citenamefont {Ko{\l}ody{\'{n}}ski},\ and\
  \citenamefont {Demkowicz-Dobrza{\'{n}}ski}}]{Haase2018}%
  \BibitemOpen
  \bibfield  {author} {\bibinfo {author} {\bibfnamefont {J.~F.}\ \bibnamefont
  {Haase}}, \bibinfo {author} {\bibfnamefont {A.}~\bibnamefont {Smirne}},
  \bibinfo {author} {\bibfnamefont {S.~F.}\ \bibnamefont {Huelga}}, \bibinfo
  {author} {\bibfnamefont {J.}~\bibnamefont {Ko{\l}ody{\'{n}}ski}}, \ and\
  \bibinfo {author} {\bibfnamefont {R.}~\bibnamefont
  {Demkowicz-Dobrza{\'{n}}ski}},\ }\bibfield  {title} {\enquote {\bibinfo
  {title} {{Precision Limits in Quantum Metrology with Open Quantum
  Systems}},}\ }\href {\doibase 10.1515/qmetro-2018-0002} {\bibfield  {journal}
  {\bibinfo  {journal} {Quantum Meas. Quantum Metrol.}\ }\textbf {\bibinfo
  {volume} {5}},\ \bibinfo {pages} {13} (\bibinfo {year}
  {2018}{\natexlab{b}})}\BibitemShut {NoStop}%
\bibitem [{\citenamefont {Ahn}\ \emph {et~al.}(2003)\citenamefont {Ahn},
  \citenamefont {Wiseman},\ and\ \citenamefont {Milburn}}]{Ahn2003}%
  \BibitemOpen
  \bibfield  {author} {\bibinfo {author} {\bibfnamefont {C.}~\bibnamefont
  {Ahn}}, \bibinfo {author} {\bibfnamefont {H.~M.}\ \bibnamefont {Wiseman}}, \
  and\ \bibinfo {author} {\bibfnamefont {G.~J.}\ \bibnamefont {Milburn}},\
  }\bibfield  {title} {\enquote {\bibinfo {title} {Quantum error correction for
  continuously detected errors},}\ }\href {\doibase 10.1103/PhysRevA.67.052310}
  {\bibfield  {journal} {\bibinfo  {journal} {Phys. Rev. A}\ }\textbf {\bibinfo
  {volume} {67}},\ \bibinfo {pages} {052310} (\bibinfo {year}
  {2003})}\BibitemShut {NoStop}%
\bibitem [{\citenamefont {Ahn}\ \emph {et~al.}(2004)\citenamefont {Ahn},
  \citenamefont {Wiseman},\ and\ \citenamefont {Jacobs}}]{Ahn2004}%
  \BibitemOpen
  \bibfield  {author} {\bibinfo {author} {\bibfnamefont {C.}~\bibnamefont
  {Ahn}}, \bibinfo {author} {\bibfnamefont {H.}~\bibnamefont {Wiseman}}, \ and\
  \bibinfo {author} {\bibfnamefont {K.}~\bibnamefont {Jacobs}},\ }\bibfield
  {title} {\enquote {\bibinfo {title} {Quantum error correction for
  continuously detected errors with any number of error channels per qubit},}\
  }\href {\doibase 10.1103/PhysRevA.70.024302} {\bibfield  {journal} {\bibinfo
  {journal} {Phys. Rev. A}\ }\textbf {\bibinfo {volume} {70}},\ \bibinfo
  {pages} {024302} (\bibinfo {year} {2004})}\BibitemShut {NoStop}%
\bibitem [{\citenamefont {Akerman}\ \emph {et~al.}(2012)\citenamefont
  {Akerman}, \citenamefont {Kotler}, \citenamefont {Glickman},\ and\
  \citenamefont {Ozeri}}]{Akerman2012}%
  \BibitemOpen
  \bibfield  {author} {\bibinfo {author} {\bibfnamefont {N.}~\bibnamefont
  {Akerman}}, \bibinfo {author} {\bibfnamefont {S.}~\bibnamefont {Kotler}},
  \bibinfo {author} {\bibfnamefont {Y.}~\bibnamefont {Glickman}}, \ and\
  \bibinfo {author} {\bibfnamefont {R.}~\bibnamefont {Ozeri}},\ }\bibfield
  {title} {\enquote {\bibinfo {title} {Reversal of photon-scattering errors in
  atomic qubits},}\ }\href {\doibase 10.1103/PhysRevLett.109.103601} {\bibfield
   {journal} {\bibinfo  {journal} {Phys. Rev. Lett.}\ }\textbf {\bibinfo
  {volume} {109}},\ \bibinfo {pages} {103601} (\bibinfo {year}
  {2012})}\BibitemShut {NoStop}%
\bibitem [{\citenamefont {Ganesan}\ and\ \citenamefont
  {Tarn}(2007)}]{Ganesan2007}%
  \BibitemOpen
  \bibfield  {author} {\bibinfo {author} {\bibfnamefont {N.}~\bibnamefont
  {Ganesan}}\ and\ \bibinfo {author} {\bibfnamefont {T.-J.}\ \bibnamefont
  {Tarn}},\ }\bibfield  {title} {\enquote {\bibinfo {title} {Decoherence
  control in open quantum systems via classical feedback},}\ }\href {\doibase
  10.1103/PhysRevA.75.032323} {\bibfield  {journal} {\bibinfo  {journal} {Phys.
  Rev. A}\ }\textbf {\bibinfo {volume} {75}},\ \bibinfo {pages} {032323}
  (\bibinfo {year} {2007})}\BibitemShut {NoStop}%
\bibitem [{\citenamefont {Szigeti}\ \emph {et~al.}(2014)\citenamefont
  {Szigeti}, \citenamefont {Carvalho}, \citenamefont {Morley},\ and\
  \citenamefont {Hush}}]{Szigeti2014}%
  \BibitemOpen
  \bibfield  {author} {\bibinfo {author} {\bibfnamefont {S.~S.}\ \bibnamefont
  {Szigeti}}, \bibinfo {author} {\bibfnamefont {A.~R.~R.}\ \bibnamefont
  {Carvalho}}, \bibinfo {author} {\bibfnamefont {J.~G.}\ \bibnamefont
  {Morley}}, \ and\ \bibinfo {author} {\bibfnamefont {M.~R.}\ \bibnamefont
  {Hush}},\ }\bibfield  {title} {\enquote {\bibinfo {title} {Ignorance is
  bliss: General and robust cancellation of decoherence via no-knowledge
  quantum feedback},}\ }\href {\doibase 10.1103/PhysRevLett.113.020407}
  {\bibfield  {journal} {\bibinfo  {journal} {Phys. Rev. Lett.}\ }\textbf
  {\bibinfo {volume} {113}},\ \bibinfo {pages} {020407} (\bibinfo {year}
  {2014})}\BibitemShut {NoStop}%
\bibitem [{\citenamefont {Genoni}\ \emph {et~al.}(2013)\citenamefont {Genoni},
  \citenamefont {Mancini},\ and\ \citenamefont {Serafini}}]{Tempura}%
  \BibitemOpen
  \bibfield  {author} {\bibinfo {author} {\bibfnamefont {M.~G.}\ \bibnamefont
  {Genoni}}, \bibinfo {author} {\bibfnamefont {S.}~\bibnamefont {Mancini}}, \
  and\ \bibinfo {author} {\bibfnamefont {A.}~\bibnamefont {Serafini}},\
  }\bibfield  {title} {\enquote {\bibinfo {title} {{Optimal feedback control of
  linear quantum systems in the presence of thermal noise}},}\ }\href {\doibase
  10.1103/PhysRevA.87.042333} {\bibfield  {journal} {\bibinfo  {journal} {Phys.
  Rev. A}\ }\textbf {\bibinfo {volume} {87}},\ \bibinfo {pages} {042333}
  (\bibinfo {year} {2013})}\BibitemShut {NoStop}%
\bibitem [{\citenamefont {Genoni}\ \emph {et~al.}(2015)\citenamefont {Genoni},
  \citenamefont {Zhang}, \citenamefont {Millen}, \citenamefont {Barker},\ and\
  \citenamefont {Serafini}}]{Levante}%
  \BibitemOpen
  \bibfield  {author} {\bibinfo {author} {\bibfnamefont {M.~G.}\ \bibnamefont
  {Genoni}}, \bibinfo {author} {\bibfnamefont {J.}~\bibnamefont {Zhang}},
  \bibinfo {author} {\bibfnamefont {J.}~\bibnamefont {Millen}}, \bibinfo
  {author} {\bibfnamefont {P.~F.}\ \bibnamefont {Barker}}, \ and\ \bibinfo
  {author} {\bibfnamefont {A.}~\bibnamefont {Serafini}},\ }\bibfield  {title}
  {\enquote {\bibinfo {title} {{Quantum cooling and squeezing of a levitating
  nanosphere via time-continuous measurements}},}\ }\href {\doibase
  10.1088/1367-2630/17/7/073019} {\bibfield  {journal} {\bibinfo  {journal}
  {New J. Phys.}\ }\textbf {\bibinfo {volume} {17}},\ \bibinfo {pages} {073019}
  (\bibinfo {year} {2015})}\BibitemShut {NoStop}%
\bibitem [{\citenamefont {Burgarth}\ \emph {et~al.}(2015)\citenamefont
  {Burgarth}, \citenamefont {Giovannetti}, \citenamefont {Kato},\ and\
  \citenamefont {Yuasa}}]{Burgarth2015a}%
  \BibitemOpen
  \bibfield  {author} {\bibinfo {author} {\bibfnamefont {D.~K.}\ \bibnamefont
  {Burgarth}}, \bibinfo {author} {\bibfnamefont {V.}~\bibnamefont
  {Giovannetti}}, \bibinfo {author} {\bibfnamefont {A.~N.}\ \bibnamefont
  {Kato}}, \ and\ \bibinfo {author} {\bibfnamefont {K.}~\bibnamefont {Yuasa}},\
  }\bibfield  {title} {\enquote {\bibinfo {title} {{Quantum estimation via
  sequential measurements}},}\ }\href {\doibase 10.1088/1367-2630/17/11/113055}
  {\bibfield  {journal} {\bibinfo  {journal} {New J. Phys.}\ }\textbf {\bibinfo
  {volume} {17}},\ \bibinfo {pages} {113055} (\bibinfo {year}
  {2015})}\BibitemShut {NoStop}%
\bibitem [{\citenamefont {Catana}\ \emph {et~al.}(2015)\citenamefont {Catana},
  \citenamefont {Bouten},\ and\ \citenamefont {Guţă}}]{Catana2015}%
  \BibitemOpen
  \bibfield  {author} {\bibinfo {author} {\bibfnamefont {C.}~\bibnamefont
  {Catana}}, \bibinfo {author} {\bibfnamefont {L.}~\bibnamefont {Bouten}}, \
  and\ \bibinfo {author} {\bibfnamefont {M.}~\bibnamefont {Guţă}},\
  }\bibfield  {title} {\enquote {\bibinfo {title} {{Fisher informations and
  local asymptotic normality for continuous-time quantum Markov processes}},}\
  }\href {\doibase 10.1088/1751-8113/48/36/365301} {\bibfield  {journal}
  {\bibinfo  {journal} {J. Phys. A}\ }\textbf {\bibinfo {volume} {48}},\
  \bibinfo {pages} {365301} (\bibinfo {year} {2015})}\BibitemShut {NoStop}%
\bibitem [{\citenamefont {Guţă}\ and\ \citenamefont
  {Kiukas}(2017)}]{Guta2016}%
  \BibitemOpen
  \bibfield  {author} {\bibinfo {author} {\bibfnamefont {M.}~\bibnamefont
  {Guţă}}\ and\ \bibinfo {author} {\bibfnamefont {J.}~\bibnamefont
  {Kiukas}},\ }\bibfield  {title} {\enquote {\bibinfo {title} {{Information
  geometry and local asymptotic normality for multi-parameter estimation of
  quantum Markov dynamics}},}\ }\href {\doibase 10.1063/1.4982958} {\bibfield
  {journal} {\bibinfo  {journal} {J. Mat. Phys.}\ }\textbf {\bibinfo {volume}
  {58}},\ \bibinfo {pages} {052201} (\bibinfo {year} {2017})}\BibitemShut
  {NoStop}%
\bibitem [{\citenamefont {Fr\"{o}wis}\ \emph {et~al.}(2014)\citenamefont
  {Fr\"{o}wis}, \citenamefont {Skotiniotis}, \citenamefont {Kraus},\ and\
  \citenamefont {D\"{u}r}}]{Frwis2014}%
  \BibitemOpen
  \bibfield  {author} {\bibinfo {author} {\bibfnamefont {F.}~\bibnamefont
  {Fr\"{o}wis}}, \bibinfo {author} {\bibfnamefont {M.}~\bibnamefont
  {Skotiniotis}}, \bibinfo {author} {\bibfnamefont {B.}~\bibnamefont {Kraus}},
  \ and\ \bibinfo {author} {\bibfnamefont {W.}~\bibnamefont {D\"{u}r}},\
  }\bibfield  {title} {\enquote {\bibinfo {title} {Optimal quantum states for
  frequency estimation},}\ }\href {\doibase 10.1088/1367-2630/16/8/083010}
  {\bibfield  {journal} {\bibinfo  {journal} {New J. Phys.}\ }\textbf {\bibinfo
  {volume} {16}},\ \bibinfo {pages} {083010} (\bibinfo {year}
  {2014})}\BibitemShut {NoStop}%
\bibitem [{\citenamefont {Demkowicz-Dobrza{\'{n}}ski}\ and\ \citenamefont
  {Maccone}(2014)}]{Demkowicz-Dobrzanski2014}%
  \BibitemOpen
  \bibfield  {author} {\bibinfo {author} {\bibfnamefont {R.}~\bibnamefont
  {Demkowicz-Dobrza{\'{n}}ski}}\ and\ \bibinfo {author} {\bibfnamefont
  {L.}~\bibnamefont {Maccone}},\ }\bibfield  {title} {\enquote {\bibinfo
  {title} {{Using Entanglement Against Noise in Quantum Metrology}},}\ }\href
  {\doibase 10.1103/PhysRevLett.113.250801} {\bibfield  {journal} {\bibinfo
  {journal} {Phys. Rev. Lett.}\ }\textbf {\bibinfo {volume} {113}},\ \bibinfo
  {pages} {250801} (\bibinfo {year} {2014})}\BibitemShut {NoStop}%
\bibitem [{\citenamefont {Huang}\ \emph {et~al.}(2018)\citenamefont {Huang},
  \citenamefont {Macchiavello},\ and\ \citenamefont {Maccone}}]{Huang2018}%
  \BibitemOpen
  \bibfield  {author} {\bibinfo {author} {\bibfnamefont {Z.}~\bibnamefont
  {Huang}}, \bibinfo {author} {\bibfnamefont {C.}~\bibnamefont {Macchiavello}},
  \ and\ \bibinfo {author} {\bibfnamefont {L.}~\bibnamefont {Maccone}},\
  }\bibfield  {title} {\enquote {\bibinfo {title} {{Noise-dependent optimal
  strategies for quantum metrology}},}\ }\href {\doibase
  10.1103/PhysRevA.97.032333} {\bibfield  {journal} {\bibinfo  {journal} {Phys.
  Rev. A}\ }\textbf {\bibinfo {volume} {97}},\ \bibinfo {pages} {032333}
  (\bibinfo {year} {2018})}\BibitemShut {NoStop}%
\bibitem [{\citenamefont {Sbroscia}\ \emph {et~al.}(2018)\citenamefont
  {Sbroscia}, \citenamefont {Gianani}, \citenamefont {Mancino}, \citenamefont
  {Roccia}, \citenamefont {Huang}, \citenamefont {Maccone}, \citenamefont
  {Macchiavello},\ and\ \citenamefont {Barbieri}}]{Sbroscia2017}%
  \BibitemOpen
  \bibfield  {author} {\bibinfo {author} {\bibfnamefont {M.}~\bibnamefont
  {Sbroscia}}, \bibinfo {author} {\bibfnamefont {I.}~\bibnamefont {Gianani}},
  \bibinfo {author} {\bibfnamefont {L.}~\bibnamefont {Mancino}}, \bibinfo
  {author} {\bibfnamefont {E.}~\bibnamefont {Roccia}}, \bibinfo {author}
  {\bibfnamefont {Z.}~\bibnamefont {Huang}}, \bibinfo {author} {\bibfnamefont
  {L.}~\bibnamefont {Maccone}}, \bibinfo {author} {\bibfnamefont
  {C.}~\bibnamefont {Macchiavello}}, \ and\ \bibinfo {author} {\bibfnamefont
  {M.}~\bibnamefont {Barbieri}},\ }\bibfield  {title} {\enquote {\bibinfo
  {title} {Experimental ancilla-assisted phase estimation in a noisy
  channel},}\ }\href {\doibase 10.1103/PhysRevA.97.032305} {\bibfield
  {journal} {\bibinfo  {journal} {Phys. Rev. A}\ }\textbf {\bibinfo {volume}
  {97}},\ \bibinfo {pages} {032305} (\bibinfo {year} {2018})}\BibitemShut
  {NoStop}%
\bibitem [{\citenamefont {Giorda}\ and\ \citenamefont
  {Allegra}(2018)}]{Giorda2016}%
  \BibitemOpen
  \bibfield  {author} {\bibinfo {author} {\bibfnamefont {P.}~\bibnamefont
  {Giorda}}\ and\ \bibinfo {author} {\bibfnamefont {M.}~\bibnamefont
  {Allegra}},\ }\bibfield  {title} {\enquote {\bibinfo {title} {Coherence in
  quantum estimation},}\ }\href {\doibase 10.1088/1751-8121/aa9808} {\bibfield
  {journal} {\bibinfo  {journal} {J. Phys. A}\ }\textbf {\bibinfo {volume}
  {51}},\ \bibinfo {pages} {025302} (\bibinfo {year} {2018})}\BibitemShut
  {NoStop}%
\bibitem [{\citenamefont {Di\'osi}\ and\ \citenamefont
  {Ferialdi}(2014)}]{Diosi2014}%
  \BibitemOpen
  \bibfield  {author} {\bibinfo {author} {\bibfnamefont {L.}~\bibnamefont
  {Di\'osi}}\ and\ \bibinfo {author} {\bibfnamefont {L.}~\bibnamefont
  {Ferialdi}},\ }\bibfield  {title} {\enquote {\bibinfo {title} {{General
  Non-Markovian Structure of Gaussian Master and Stochastic Schr\"odinger
  Equations}},}\ }\href {\doibase 10.1103/PhysRevLett.113.200403} {\bibfield
  {journal} {\bibinfo  {journal} {Phys. Rev. Lett.}\ }\textbf {\bibinfo
  {volume} {113}},\ \bibinfo {pages} {200403} (\bibinfo {year}
  {2014})}\BibitemShut {NoStop}%
\end{thebibliography}%

\end{document}